 \documentclass[final,5p,times,twocolumn]{elsarticle}

 \usepackage{graphics}

 \usepackage{amsmath}

\begin{document}

\begin{frontmatter}



\title{Commissioning of the BRIKEN detector for the measurement of very exotic $\beta$-delayed neutron emitters}


\author[ific]{A. Tolosa-Delgado}
\author[ific]{J. Agramunt}
\author[ific]{J. L. Tain\corref{cor1}}
\cortext[cor1]{Tel.: +34 963543497, Fax: +34 963543488.
Instituto de F\'{\i}sica Corpuscular, Catedr\'{a}tico Jos\'{e} Beltr\'{a}n 2, 
E-46980 Paterna, Spain}
\ead{tain@ific.uv.es}
\author[ific,atomki]{A. Algora}
\author[ific]{C. Domingo-Pardo}
\author[ific]{A. I. Morales}
\author[ific]{B. Rubio}

\author[upc]{A. Tarife\~{n}o-Saldivia}
\author[upc]{F. Calvi\~{n}o}
\author[upc]{G. Cortes}

\author[ornl]{N. T. Brewer}
\author[ornl]{B. C. Rasco} 
\author[ornl]{K. P. Rykaczewski} 
\author[ornl]{D. W. Stracener} 
\author[ornl]{J. M. Allmond}

\author[utk]{R. Grzywacz} 
\author[utk]{R. Yokoyama} 
\author[utk]{M. Singh} 
\author[utk]{T. King} 
\author[utk]{M. Madurga}

\author[riken]{S. Nishimura} 
\author[riken,vast]{V.H. Phong} 
\author[riken]{S. Go} 
\author[riken,uhk]{J. Liu} 
\author[riken,utokyo]{K. Matsui} 
\author[riken,utokyo]{H. Sakurai} 
\author[riken,atomki]{G. G. Kiss}
\author[riken]{T. Isobe} 
\author[riken]{H. Baba} 
\author[riken]{S. Kubono} 
\author[riken]{N. Fukuda} 
\author[riken]{D.S. Ahn} 
\author[riken]{Y. Shimizu} 
\author[riken]{T. Sumikama}
\author[riken]{H. Suzuki}
\author[riken]{H. Takeda}
\author[riken]{P. A. S\"oderstr\"om}

\author[niigata]{M. Takechi}

\author[edinburgh]{C. G. Bruno}
\author[edinburgh]{T. Davinson}
\author[edinburgh]{C. J. Griffin}
\author[edinburgh]{O. Hall}
\author[edinburgh]{D. Kahl} 
\author[edinburgh]{P. J. Woods}

\author[daresbury]{P. J. Coleman-Smith} 
\author[daresbury]{M. Labiche} 
\author[daresbury]{I. Lazarus} 
\author[daresbury]{P. Morrall} 
\author[daresbury]{V. F. E. Pucknell} 
\author[daresbury]{J. Simpson}

\author[rutherford]{S. L. Thomas} 
\author[rutherford]{M. Prydderch}

\author[liverpool]{L. J. Harkness-Brennan} 
\author[liverpool]{R. D. Page}

\author[triumf]{I. Dillmann} 
\author[triumf]{R. Caballero-Folch}
\author[triumf]{Y. Saito}

\author[cmu]{A. Estrade} 
\author[cmu]{N. Nepal}

\author[nscl]{F. Montes}

\author[npl,surrey,riken]{G. Lorusso}

\author[mcmaster]{J. Liang}

\author[snu]{S. Bae}

\author[snu,riken]{J. Ha}

\author[ku]{B. Moon}

\author{the BRIKEN collaboration}


\address[ific]{Instituto de F\'{\i}sica Corpuscular, 
CSIC and Universitat de Valencia, E-46980 Paterna, Spain}

\address[upc]{Universitat Politecnica de Catalunya, 
E-08028 Barcelona, Spain}

\address[ornl]{Oak Ridge National Laboratory, 
Physics Division, TN 37831-6371, USA}

\address[utk]{University of Tennessee, 
 Knoxville, Tennessee, USA}

\address[riken]{RIKEN Nishina Center, 
Wako, Saitama 351-0198, Japan}

\address[utokyo]{University of Tokyo, 
Department of Physics, Tokyo 113-0033, Japan}

\address[niigata]{Graduate School of Science and Technology,
Niigata University, Niigata 950-2102, Japan}

\address[edinburgh]{University of Edinburgh,
School of Physics and Astronomy, Edinburgh EH9 3FD, UK}

\address[daresbury]{STFC Daresbury Laboratory, 
Daresbury, Warrington WA4 4AD, UK}

\address[rutherford]{STFC Rutherford Appleton Laboratory, 
Harwell Campus, Didcot OX11 0QX, UK}

\address[liverpool]{Department of Physics,
University of Liverpool,  Liverpool L69 7ZE, UK}

\address[triumf]{TRIUMF, 
Vancouver BC, V6T 2A3, Canada}

\address[cmu]{Central Michigan University, 
Mount Pleasant, MI 48859, USA}

\address[nscl]{National Superconducting Cyclotron Laboratory, 
East Lansing, Michigan 48824, USA}

\address[npl]{National Physical Laboratory
Teddington, TW11 0LW, UK}

\address[surrey]{University of Surrey,
Department of Physics, Guildford, GU2 7XH,  UK}

\address[atomki]{MTA Atomki, 
Debrecen, H4032, Hungary}

\address[vast]{Institute of Physics, 
Vietnam Academy of Science and Technology,  Hanoi, Vietnam}

\address[uhk]{University of Hong Kong, 
Department of Physics, Pokfulman Road, Hong Kong}

\address[mcmaster]{McMaster University, 
Department of Physics and Astronomy, Hamilton, Ontario L8S 4M1, Canada}

\address[snu]{Seoul National University,
Department of Physics and Astronomy,  Seoul 08826, Republic of Korea}

\address[ku]{Korea University, 
Department of Physics, Seoul 136-701, Republic of Korea}

\begin{abstract}
A new detection system  has been installed at the RIKEN Nishina Center (Japan)
to investigate decay properties of very neutron-rich nuclei.
The setup consists of three main parts: a moderated neutron counter, 
a detection system sensitive to the
implantation and decay of radioactive ions, and $\gamma$-ray detectors.
We describe here the setup, the commissioning experiment and some selected results
demonstrating its performance for the measurement of half-lives and $\beta$-delayed
neutron emission probabilities. The methodology followed in the analysis of the data is 
described in detail. Particular emphasis is placed on the correction of the accidental 
neutron background.
\end{abstract}


\end{frontmatter}


\section{Introduction}
\label{introduction}
$\beta$-delayed neutron decay is a rare process on Earth, happening in nuclear power reactors,
but it dominates the disintegration of nuclei produced during the
rapid (r) neutron capture process in explosive stellar events \cite{kod73}. 
In such environments, an intense burst of neutrons
synthesizes, in a short time, very neutron-rich unstable nuclei
for which the neutron separation energy $S_{1n}$ in the daughter is smaller than
the decay energy window $Q_{\beta}$. 
It can happen that also the two-neutron separation energy $S_{2n}$, 
in general the $x$-neutron separation energy $S_{xn}$, is smaller than
$Q_{\beta}$ leading to multiple neutron emission. 
The decay energy window for $xn$ emission is defined as $Q_{\beta xn} = Q_{\beta} - S_{xn}$.
The branchings for this decay mode and the number of neutrons emitted
per decay are important quantities for our understanding of the abundance of
stable elements produced at the end of the decay chain
following neutron exhaustion in the r-process.
The probability for the emission of $x$ neutrons is designated as $P_{xn}$ 
and the total neutron emission probability is $P_{n} = \sum_{x=1}^{max} P_{xn}$. 
The probability of decay with no-neutron emission is just $P_{0n} = 1-P_{n}$.
The average number of neutrons per decay, or neutron multiplicity,
is $M_{n} = \sum_{x=1}^{max} x P_{xn}$. 
Another quantity of key astrophysical interest is the decay half-life $T_{1/2}$ 
of the nuclei along the path of nucleosynthesis, 
governing the initial abundances and the speed of the r-process.

Determining experimentally $P_{xn}$ and $T_{1/2}$
values for very exotic nuclei is one of the 
goals of current research in nuclear astrophysics \cite{mum16}. The challenges are to
produce with sufficient intensity the relevant nuclei located
far from the valley of $\beta$-stability
and to measure accurately the corresponding quantities in their decay.
The BRIKEN collaboration \cite{tai18} aims to expand
our current knowledge \cite{crpdn} on $P_{xn}$ and $T_{1/2}$ values 
to the most exotic neutron-rich nuclei that are accessible.
To achieve this,  advanced instrumentation has been developed to be used 
at state-of-the-art radioactive beam facilities. Our approach to the measurement
of $P_{xn}$ is to use direct neutron counting to select the $\beta xn$
channel in combination with $\beta$ counting which provides the
total number of decays. A new high efficiency neutron counter has been 
designed \cite{tar17} and assembled for this purpose.
From the different detector configurations studied 
in \cite{tar17} we chose the one
including two CLOVER-type HPGe detectors, for $\gamma$ spectroscopy, 
that maximizes the total neutron detection efficiency $\varepsilon_{n}$
and at the same time
minimizes the dependence of $\varepsilon_{n}$ on neutron energy $E_{n}$
in the 0-5~MeV range.
The detector was combined with the  Advanced Implantation and Decay Array (AIDA)
\cite{gri17} and installed at the RIKEN Nishina Center.
The setup was commissioned with radioactive beams in a parasitic run in November 2016 
using neutron-rich nuclei around mass number $\mathrm{A}=80$.
The first experimental campaign took place in May-June 2017
with measurements on nuclei with $\mathrm{A} \sim 80$, $\mathrm{A} \sim 130$ 
and $\mathrm{A} \sim 160$. The second campaign in October-November 2017
collected data for $\mathrm{A} \sim 80$ and $\mathrm{A} \sim 100$.
New experiments in other mass regions are planned.

This publication focuses on data from the commissioning run.
The setup and the
measurements are described in Section \ref{sec:experiment}.
Section \ref{sec:analysis} describes the methodology followed in the analysis
of data specific to this type of experiments. 
The accurate background correction of the data 
turns out to be critical and a novel method is described
in Section \ref{sec:background}. Some selected results showing the 
performance of the setup are presented in Section
\ref{sec:results}.

\section{Experimental details}
\label{sec:experiment}

\begin{figure}[ht]
 \begin{center}
 \includegraphics[width=7.8cm]{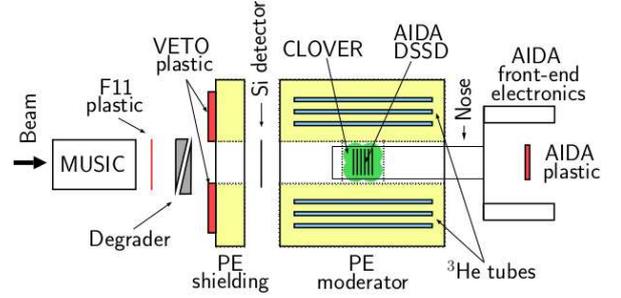}
 \end{center}
 \caption{Arrangement of elements in the experimental setup mentioned in the text. The drawing is not to scale.}
 \label{fig:sketch}
\end{figure}

A schematic drawing of the disposition of different elements described below, 
belonging to the experimental setup  at the end of the beam line,
is shown in Fig.~\ref{fig:sketch}.

\subsection{Measurements}
\label{sec:measurement}
\begin{figure}[ht]
 \begin{center}
 \includegraphics[width=7.8cm]{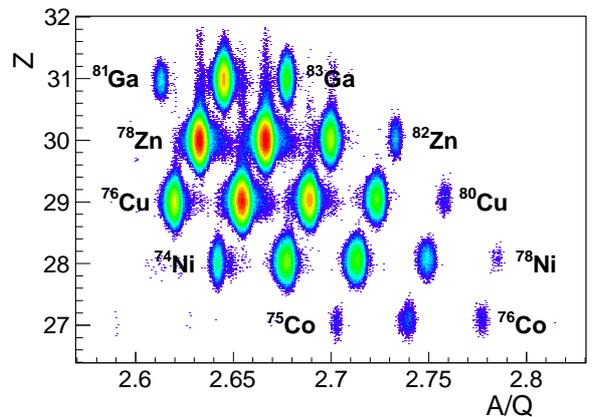}
 \end{center}
 \caption{Identification plot of ions implanted in AIDA during the commissioning run. 
The plot shows the atomic number Z versus the mass-to-charge ratio A/Q of the ion.
The setting of the BigRIPS spectrometer was centered on $^{76}$Ni.}
 \label{fig:pid}
\end{figure}

The experiments were performed using primary beams of $^{238}$U at
high intensity (20-50~pnA) accelerated to an energy of 345~MeV per nucleon
by the accelerator complex of the Radioactive Isotope Beam Factory (RIBF)  \cite{oku12}.
The beam hits a beryllium target 4~mm thick producing a large number
of fast reaction products which are selected by the BigRIPS in-flight separator and guided
to the F11 experimental area through the Zero-Degree Spectrometer (ZDS) \cite{kub12}.
Each ion in the cocktail of nuclei arriving at the measuring station
is identified though measurement of 1) its atomic charge Z, 
and 2) its mass-to-charge ratio A/Q. 
These quantities are obtained from the magnetic 
rigidity B$\rho$, the time-of-flight (ToF)  and the energy loss ($\Delta E$) of the ion.
This information is provided by the spectrometer 
and its ancillary detectors: plastic scintillation detectors, position-sensitive parallel plate 
avalanche counters (PPAC) and multi-sampling ionization chambers (MUSIC). 
The last elements of the beam line  were a pair of MUSIC
detectors and a thin (1~mm thick) plastic scintillation detector (F11 plastic) 
with an area of $12~\mathrm{cm} \times 10~\mathrm{cm}$ (see Fig.~\ref{fig:sketch}).
The ions of interest were implanted in the AIDA detector
adjusting their velocity by means of an aluminum  degrader of variable thickness
situated after the F11 plastic.
An identification plot of the ions implanted in AIDA during the commissioning run 
is shown in Fig.~\ref{fig:pid}. 
The setting of the BigRIPS spectrometer was centered on $^{76}$Ni.
Neutron-rich isotopes from cobalt to gallium 
were implanted, most of which are $\beta$-delayed neutron emitters.
This included 475 events for the doubly magic $^{78}$Ni.

\subsection{Setup and instrumentation}
\label{sec:setup}

The implantation detector AIDA \cite{aida} consists of a stack of six silicon 
double-sided strip detectors (DSSD) with a spacing of 10~mm between them. 
The PCB frame of the DSSD is suspended at the corners on thin titanium rods 
inside the AIDA nose made of 1~mm thick aluminum with a cross-section of 
$10~\mathrm{cm} \times 10~\mathrm{cm}$. 
The nose is closed on the beam side by an aluminized Mylar
foil to ensure light tightness.
Each DSSD has a thickness of 1~mm and an area 
of $71.68~\mathrm{mm} \times 71.68~\mathrm{mm}$, 
with 128 strips 0.51~mm wide on each side. The strips on the two sides 
are perpendicular to each other
and provide high resolution position information 
in the horizontal (X) and vertical (Y) directions.
Specially made flat cables running inside the nose bring the strip signals  to the
front-end readout electronics located about 70~cm away.
Dual electronic chains are used to process the signals from each strip. The low-gain
branch (20 GeV range) is used to process the high energy implantation signals.
The high-gain branch (20 MeV range) is used to identify the much lower energy signals
from $\beta$ particles emitted in the decay of the radioactive ions.
The dual electronic processing, implemented using ASICs, 
minimizes the overload recovery time for $\beta$ registration to a few $\mu$s.
The stack of Si DSSDs is positioned at the geometrical centre of the neutron detector
with the electronics located outside, downstream in the beam direction.
A plastic scintillation detector of thickness 10~mm (AIDA plastic)  is positioned
on the beam axis 120~cm downstream of the stack
to detect particles that pass through.

During the experiments in October-November 2017 the AIDA detector was replaced by
the WA3ABi detector \cite{nis13} which consists of a stack of four Si DSSDs with 3~mm wide strips.
The advantage is that these DSSDs are narrower ($50~\mathrm{mm} \times 50~\mathrm{mm}$)
allowing us to move the CLOVER detectors closer and increase the $\gamma$ 
detection efficiency. In addition we used a new implantation-decay detector made
of YSO scintillation material developed at the University of Tennessee \cite{grz18,als15}. 
This detector consists of an array of 
$48 \times 48$ closely packed crystals with dimension 
$1~\mathrm{mm} \times 1~\mathrm{mm} \times 5~\mathrm{mm}$. The array is
coupled through a light guide to a H8500B flat panel type photomultiplier tube (PMT) 
with an $8 \times 8$
segmented anode that is readout with a resistor network. 
Both detectors were used at the same time with WA3ABi positioned
off-centre $\sim 20$~mm upstream and the 
YSO detector positioned off-centre $\sim 20$~mm downstream.
A full description of this implantation-decay setup and its performance
will be given in a forthcoming publication.

During the May-June 2017 experiments we added a  thin large area Si detector to the setup.
The purpose of this $\Delta E$ detector is to help in the identification of light particles
(p, d, $\alpha$, ...) coming with the beam. It is a 
single sided strip detector of quasi-rectangular shape and
dimension $134~\mathrm{mm} \times 123~\mathrm{mm}$ with a thickness of 330~$\mu$m.
It has 26 horizontal strips combined into two readout channels (top and bottom).
The detector was placed about 50~cm upstream before the neutron detector.

\begin{figure}[ht]
 \begin{center}
 \includegraphics[width=6.cm]{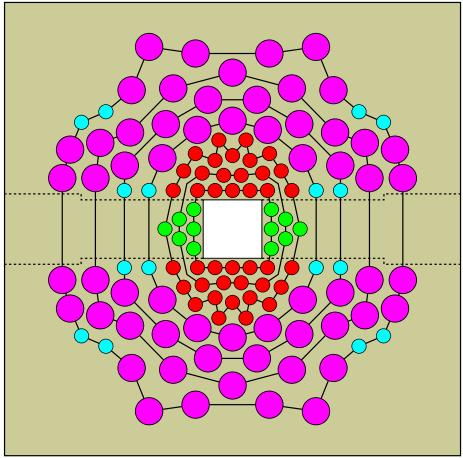}
 \end{center}
 \caption{Distribution of $^{3}$He tubes around the AIDA hole (in white). 
The size of the transverse holes for the CLOVER detectors is indicated
with the dashed line.
The color indicates the type of tube
 (see Table~\ref{tab:tubes}). Green: RIKEN; red: UPC; light-blue: ORNL1; pink: ORNL2.
The black continuous line connects the tubes belonging to each of the seven rings defined.
Ring 1 is the inner most. Ring 7 is the outer most.}
 \label{fig:tubes}
\end{figure}

The BRIKEN neutron counter consists of an array of 140 $^{3}$He filled proportional tubes
embedded in a large volume of polyethylene (PE) acting as a neutron energy moderator.
Very low-energy neutrons have a large interaction probability with the gas in the tubes
through the reaction $\mathrm{n} + \mathrm{^{3}He} \rightarrow  \mathrm{^{3}H} + \mathrm{p}$. 
This reaction liberates an energy of 764~keV that is easily detected.
The PE moderator has external dimensions of 
$90~\mathrm{cm} \times 90~\mathrm{cm} \times 75~\mathrm{cm}$, 
with a longitudinal hole (in the beam direction) of cross-section 
$11.6~\mathrm{cm} \times 11.6~\mathrm{cm}$
into which AIDA is inserted from the back. 
The PE moderator is constructed as a stack of 5~cm thick slabs in the longitudinal direction
held together by stainless steel rods passing through the corners.
The lateral sides and the top of the PE volume are covered with
1~mm thick Cd sheets  and additional slabs of PE of 25~mm
for neutron background attenuation.
The two CLOVER detectors are inserted 
horizontally from opposite sides 
into transverse holes of cross-section 
$11~\mathrm{cm} \times 11.6~\mathrm{cm}$
facing the stack of DSSDs.
Four different types of $^{3}$He tube were used in the
array and their characteristics are summarized in Table~\ref{tab:tubes}.
The UPC tubes come from the BELEN detector \cite{gom11} and the ORNL tubes
come from the 3Hen detector \cite{grz14}.
The RIKEN and ORNL tubes were manufactured by GE Reuter Stokes \cite{GE}
and the UPC tubes by LND Inc \cite{LND}.
The 60~cm long UPC, ORNL1 and ORNL2 tubes are arranged around the AIDA hole
and are centred longitudinally on the DSSD stack.
The shorter RIKEN tubes (30~cm) are disposed on both sides 
of each CLOVER detector hole. The transverse position distribution
of the tubes is symmetrical and follows an approximate ring geometry 
as indicated in Fig.~\ref{fig:tubes}. 

\begin{figure}[ht]
 \begin{center}
 \includegraphics[width=7.8cm]{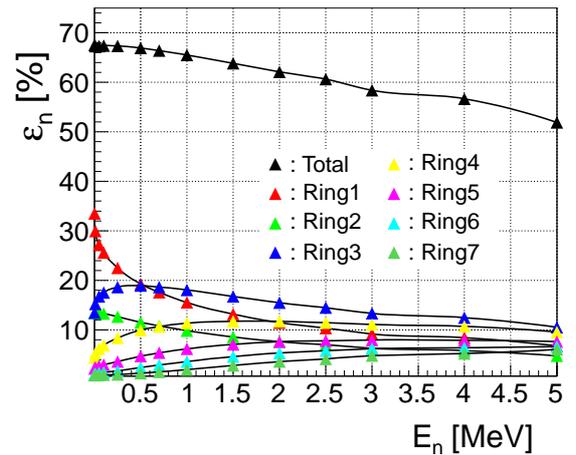}
 \end{center}
 \caption{Efficiency of the BRIKEN neutron counter as a function of neutron energy.
The total efficiency (black symbols) and the contribution of each ring (colored symbols)
is shown. See Fig.~\ref{fig:tubes} for the definition of rings.}
 \label{fig:neff}
\end{figure}

We have calculated the efficiency of the neutron detector with 
Monte Carlo (MC) simulations using the Geant4 Simulation Toolkit \cite{geant4}. 
Figure \ref{fig:neff} shows
the total efficiency and the efficiency per ring as a function of neutron energy.
Up to 0.5~MeV the total efficiency varies within $\pm 0.3 \%$ and has an
average value of 67.2\%. The efficiency decreases to 65.5\% at 1~MeV,
then drops to 60.6\% at 2.5~MeV and 51.9\% at 5~MeV.
We used experimental neutron spectra \cite{bra89,endfb71} 
to simulate average efficiencies for a few known $\beta$-delayed neutron emitters
with $Q_{\beta 1n}$ values between
2~MeV and 5.8~MeV. The resulting efficiencies vary from 67.2\% to 66.1\%.
A similar simulation was made using the known spectrum of $^{252}$Cf \cite{man87}
extending up to 20~MeV  with an average energy of $\langle E_{n} \rangle = 2.13$~MeV 
and a value of 61.8\% was obtained . This value agrees well with the experimental  result of
61.4(17)\% obtained during the characterization of the BRIKEN neutron counter 
with  a $^{252}$Cf source \cite{tar18}. From these results we set the nominal
neutron detection efficiency of the counter in the present configuration for
isotopes with low or moderate $Q_{\beta 1n}$ windows to 
$\bar{\varepsilon}_{n} = 66.8 (20)\%$. This value is further investigated below
(Section \ref{sec:results})
using the results of measurements presented here.

\begin{table}[ht]
\caption{Main characteristics (gas volume and pressure) and number
of the different types of $^{3}$He tubes used in the BRIKEN neutron counter.}
\begin{center}
\begin{tabular}{ccccc} \hline
Type & Length & Diameter & Pressure  & Number\\
	& (mm) & (mm) & (atm)  & \\ \hline
RIKEN & 300 & 25.4 & 5 & 24 \\
UPC & 600 & 25.4 & 8  & 40 \\ 
ORNL1 & 609.6 & 25.4 & 10 & 16 \\ 
ORNL2 & 609.6 & 50.8 & 10 & 60 \\ \hline
\end{tabular}
\end{center}
\label{tab:tubes}
\end{table}

We placed a PE shielding against fast neutrons coming from the beam line
approximately 60~cm upstream of the neutron detector.
The shielding has a thickness of 20~cm, a cross-section of 
$90~\mathrm{cm} \times 90~\mathrm{cm}$ and a central
hole for the beam of $11.6~\mathrm{cm} \times 11.6~\mathrm{cm}$.
Cadmium sheets were fastened to the back of the shielding.
At the front of the shielding two large plastic scintillation detectors were attached,
above and below the hole, with dimensions 
$45~\mathrm{cm} \times 20~\mathrm{cm} \times 1~\mathrm{cm}$.
These detectors serve to discriminate against fast neutrons from the beam (VETO plastics).
 
The $^{3}$He tubes are connected to the preamplifiers via double-shielded coaxial cables
to minimize noise pickup. 
These are Mesytec MPR-16-HV modules with 16 independent channels \cite{mpr16}.
A total of 10 modules are used to accommodate all the tubes. Four of them have
a differential output and the remainder have unipolar outputs. Before being sent to the
sampling digitizer modules, the differential
signals are converted into unipolar signals using 16 channel converter cards
designed at the Accelerator Laboratory of the University of Jyv\"{a}skyl\"{a} (JYFL). 
A common high voltage (HV) is applied to all the tubes connected to a preamplifier module
using a remotely controllable MPOD system from Wiener with ISEG HV cards \cite{mpod}.  
The slow control system for this and other ancillary instrumentation was developed
at ORNL (C. J. Gross and N. T. Brewer).
The voltages applied are
1450~V for RIKEN and UPC tubes, 1350~V for  ORNL1 tubes , and 1750~V for ORNL2 tubes.
A common pulser signal is fed to all preamplifier modules. The pulse
generator is driven by a precision clock running at 10~Hz.
One of the pulser signals is sent directly to a free digitizer channel.
The pulser is used to determine the data acquisition live time accurately.

The CLOVER detectors come from the CLARION array of 
Oak Ridge National Laboratory \cite{gro00}.
The four crystals in each detector have a diameter of 50~mm and a length around 80~mm.
They are assembled inside the Al nose at 10~mm from the front face.
The nose has a section of $10.1~\mathrm{cm} \times 10.1~\mathrm{cm}$.
We use the preamplified signals from the central contacts (eight in total)
which are sent directly to a digitizer module. The HV is provided by the MPOD system.

\begin{figure}[ht]
 \begin{center}
 \includegraphics[width=7.8cm]{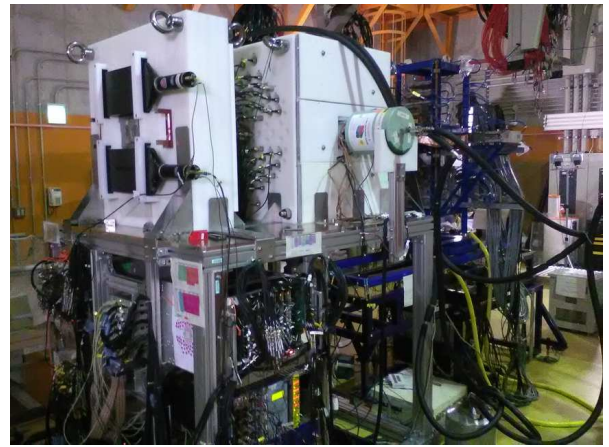}
 \end{center}
 \caption{Photograph of the full BRIKEN setup used during the first measurements.
The beam is coming from the left.
At the front-left side is the PE shielding for beam neutrons with the two plastic veto detectors.
At the center behind the PE shielding is the BRIKEN neutron detector.
Also visible is the dewar of one CLOVER detector inserted in the neutron detector  PE moderator. 
At the right-back side of the figure is visible the structure supporting the AIDA front-end electronics.
The electronics and DACQ for BRIKEN is located below the detector.}
 \label{fig:setup}
\end{figure}

A picture of the full setup can be seen in  Fig.~\ref{fig:setup}. 

\subsection{Data acquisition and sorting}
\label{sec:dacq}
Both the AIDA detector and the BigRIPS spectrometer have their proprietary
data acquisition systems (DACQ). 

For the BRIKEN neutron counter we used the self-triggered
 Gasific DACQ developed at IFIC (Valencia)
\cite{agr16}. An upgrade was needed in order to handle the large number of electronic channels.
The new system uses two VME crates to accommodate seven SIS3316 and seven SIS3302
sampling digitizers from Struck \cite{struck}. 
The SIS3316 features 16 digitizer channels, with 14 bits
and a maximum sampling rate of 250~MS/s, while the SIS3302 has 8 digitizers channels with
16 bits and a maximum 100~MS/s sampling rate. For each channel 
amplitude and time information are registered for signals above a specified
threshold. A common clock distributor
SIS3820 is used to synchronize the sampling in all the modules. The  clock
frequency was set to 50~MHz. It is also possible to run some of the digitizers
at a multiple of that frequency using a special feature of the firmware,
which is an advantage when combining fast and slow detectors.
The Gasific DACQ also handles the electronic pulses from the CLOVER
detectors and other ancillary detectors such as the F11 plastic detector, the AIDA plastic detector,
the VETO plastic detectors and the Si $\Delta E$ detector. 
It was also used to acquire the fast signals from the YSO detector at 250~MS/s,
using the upscaling of the sampling rate feature in the DACQ.
The signals from the fast plastic detectors were shaped before entering the digitizer.
The digitized signals are processed on-board with a fast trapezoidal filter providing
noise discrimination and timing information. Accepted signals are
timestamped and processed with a trapezoidal filter with compensation for the
preamplifier decay constant to obtain 
the amplitude (energy) information.
The parameters of both digital filters are optimized for every detector type.
Parameter setting, acquisition control and on-line data surveillance is performed by Gasific.

To perform a complete data analysis it is necessary to combine the information from the
three independent DACQs: BigRIPS, AIDA and BRIKEN.  This is done on the basis
of the absolute time-stamps, thanks to the use of a
common synchronization signal distributed to all three systems.
Since maintaining the synchronization is crucial for the success of the measurement,
we developed an on-line monitoring program that periodically spies on the
timestamps on the three data streams and checks that the events are synchronized.

\begin{figure}[ht]
 \begin{center}
 \includegraphics[width=6.8cm]{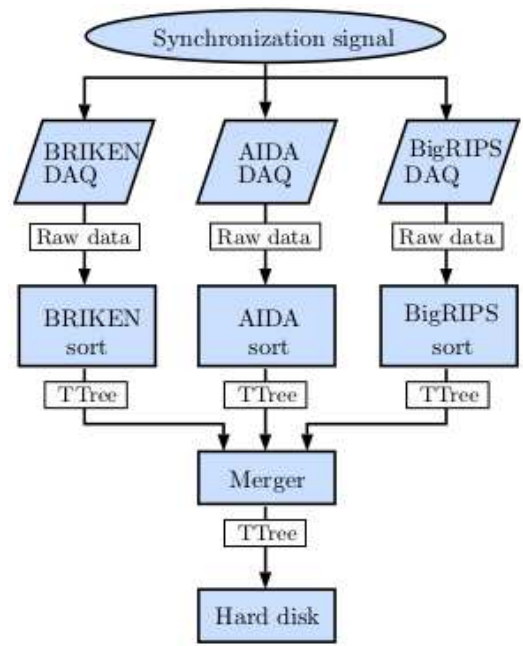}
 \end{center}
 \caption{Diagram of the process for data merging. See text for details.}
 \label{fig:merger}
\end{figure}

We developed an efficient scheme for data processing which gives us 
the possibility of performing a detailed off-line analysis with information 
from the three systems within a few hours (near-line analysis). 
This allows us to asses
the progress of the measurement and to detect experimental issues
that need corrective action. The scheme is shown in Fig.~\ref{fig:merger}.
A new run is started every hour and the data from the previous run is
copied to a dedicated server. The raw data from every detector system are then processed 
with a specific sorting program which generates a ROOT TTree \cite{root} file
from each data stream. These TTrees contain for 
each event type the necessary information.
The minimum information required,
apart from the time-stamp, consists of: 1)  BigRIPS: the Z and A/Q of each ion,
2) AIDA: the X, Y, Z position and the energy E of each ion or $\beta$ signal, and
3) BRIKEN: a detector identifier and the energy E of each signal.

To combine the information of the three TTrees in a single TTree
a \emph{Merger} software program has been developed.
The program uses C++ containers to efficiently merge and order the data by time.
It can also associate ROOT vectors with each output event,  containing presorted
time ordered data of different event types. This boosts the construction of 
time correlations in the off-line analysis. For example, each $\beta$ event can have
a vector of implant events and a vector of neutron events occurring within 
specified time ranges around the $\beta$ event.

\subsection{Detector performance}
\label{sec:performance}
\begin{figure}[ht]
 \begin{center}
 \includegraphics[width=7.8cm]{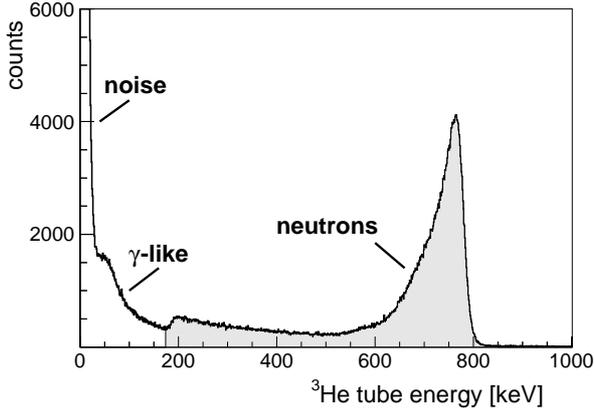}
 \end{center}
 \caption{Distribution of energies registered in the $^{3}$He tubes
of the BRIKEN neutron counter.
The spectrum contains the sum of all $^{3}$He tube signals in one run.
The shaded area represents the range of signals accepted as valid neutron signals.
See text for details.}
 \label{fig:en3he}
\end{figure}

Figure \ref{fig:en3he} shows the energy spectrum registered in the $^{3}$He tubes
during one run with beam. 
The shape of the tube response to neutrons, represented by the shadowed 
area, is the sum of all tube dependent responses.
The characteristic 
full absorption peak at 764~keV serves to calibrate in energy all the tubes at the beginning
of the run. In general the gain of the tubes is very stable during
the measurement, with occasional minor jumps for some of them
that do not even require a gain correction. 
The resolution and the tail produced 
by the wall effect determine the range of signals identified
as neutrons: 175~keV - 850~keV. 
The data represented in the spectrum of Fig.~\ref{fig:en3he} was taken with a very
low acquisition threshold.
The peak observed below 30~keV is dominated by electronic noise.
Above 30~keV another component is seen, that we call $\gamma$-like.
We associate this component with radiation induced by the beam on different 
material elements in its path close to the neutron detector. 
It extends well into the neutron signal range, 
thereby contributing to the accidental neutron background.
See also the discussion related to Fig.~\ref{fig:tf11n} below.
We found that the LND tubes are less sensitive than GE
tubes to this background contribution which otherwise shows a radial
intensity profile decreasing with distance from the beam axis.

\begin{figure}[ht]
 \begin{center}
 \includegraphics[width=7.8cm]{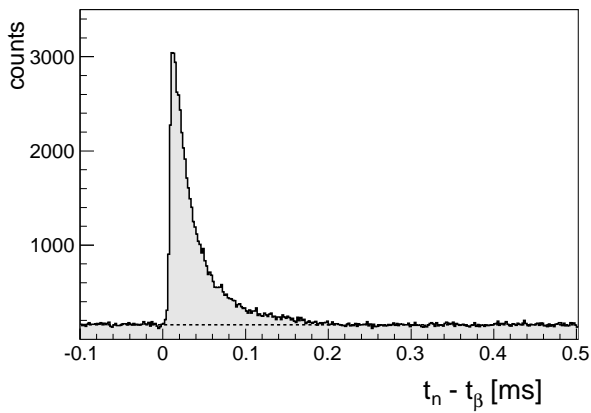}
 \end{center}
 \caption{Distribution of time differences between neutron signals in BRIKEN 
and $\beta$ signals in AIDA ($t = t_{n} -t_{\beta}$)
showing the neutron moderation plus capture time distribution. 
The flat contribution represented with a dashed line 
corresponds to accidental correlations.}
 \label{fig:tmod}
\end{figure}

The neutron energy moderation process plus the time needed for
a thermalized neutron to be absorbed in a $^{3}$He tube introduce 
a considerable delay between neutron production and its detection.
Figure \ref{fig:tmod} shows the time distribution between 
neutron signals in the whole BRIKEN detector and signals 
identified as $\beta$ particles in AIDA. The tail of the distribution
shows more than one exponential component
but is essentially contained within the interval of 200~$\mu$s (99.6\%).
Compared with other neutron counters of the same kind
(see for example Ref.~\cite{gom11})  this
distribution is rather short. This is a consequence of
the close packing of tubes in our arrangement.
Based on the moderation plus capture time spectrum we decided to
use a $\beta$-neutron coincidence time window of $\Delta t_{\beta n} = 200$~$\mu$s  
to correlate neutrons with decays.  Shorter windows
could be used to reduce the ratio of accidentally correlated neutrons,
represented by the flat background in Fig. \ref{fig:tmod},
at the price of reducing the neutron detection efficiency.

The long coincidence window of 200~$\mu$s  would introduce unnecessary common dead-time 
in a triggered event based DACQ. This is the main reason to use
our DACQ where every individual channel runs in self-triggered mode.
We determine the life-time for every channel as the ratio of counts
in the pulser peak appearing in the $^{3}$He energy spectrum 
(located outside the range shown in Fig.~\ref{fig:en3he})
to the pulser counts registered in an independent channel where the pulser signals
are directly connected. 
In general we observe a very small dead-time.
For example during a measurement with
a $^{252}$Cf source the total rate in BRIKEN was 15730~cps,
including noise. The rate per tube varies strongly depending
on the position of the tube. The highest channel rate amounted to 255~cps
and the lowest to 11~cps. The measured channel dead-time fractions
were 0.40\% and 0.01\% respectively. These numbers agree well with
the estimation for a non-paralyzable system using the 
channel trigger gate length set via software. The global dead-time is 0.36\%,
obtained by weighting the channel dead-times with their relative contribution 
to the total number of counts. 
In comparison, an event-based
DACQ with a 200~$\mu$s gate will have a dead-time of 76\% at a rate of 15.7~kcps.
During the experimental runs the rate in BRIKEN was never
higher than a few hundred counts-per-second thus the acquisition 
dead-time corrections are negligible ($<0.1$\%).

One of the issues encountered during the commissioning run was the large
rate of beam induced neutrons, 
dominating the neutron background in BRIKEN.
This came as no surprise since in a previous experiment \cite{cab17} with the BELEN
neutron detector at the GSI Fragment Separator (FRS) we observed in some cases
more than 250 neutrons/s. 
The large background rate is a consequence of the high energy of the
radioactive beam. 
We found the neutron rate at BigRIPS to be sensitive to the spectrometer setting
and to the amount of material in the beam path, in particular
close to the detector. Whenever possible we tried to move the material away from the
experimental area. For example reducing the secondary beam energy 
in the early stages of the spectrometer
allows us to reduce the thickness of the variable degrader controlling the
implantation. In spite of these measures the observed rate is still large.
During the commissioning run we measured
up to 200 neutrons/s and in later experiments up to 160 neutrons/s.
For comparison the rate induced by the natural background 
is 0.4~neutrons/s. This quite low rate is a consequence
of the location of the experimental area, around 20~m underground.

We observed that a large fraction of background neutrons is time correlated with 
the signals of ions passing through the F11 plastic. Figure \ref{fig:tf11n}
shows the correlation time distribution, where the characteristic neutron
moderation curve is seen. When comparing this figure with Fig. \ref{fig:tmod}
two differences can be seen: the spike at $t=0$ and the longer tail
of the distribution. The spike is due to $\gamma$-like signals
within the neutron signal range. This is demonstrated by the grey filled
histogram in Fig.~\ref{fig:en3he} obtained gating on signals in the tubes above the noise
but below 165~keV. 
The longer moderation time observed in Fig.~\ref{fig:tf11n}, up to 500~$\mu$s, 
is likely to be the consequence
of the high energy and direction of incidence of beam background neutrons. 

We exploited this correlation to reduce the neutron background effect
by imposing off-line a veto condition whenever a neutron is preceded a short time before
by an ion signal in the F11 plastic. This veto condition introduces
an analysis dead-time that is proportional to the rate in the F11 plastic.
During the commissioning run the rate in the F11 plastic was 460~cps in average,
thus we decided to use a veto time window of $\Delta t_{F11n} = 200$~$\mu$s
which captures 96.4\% of background signals and gives a veto
dead-time of 8.79\%.

\begin{figure}[ht]
 \begin{center}
 \includegraphics[width=7.8cm]{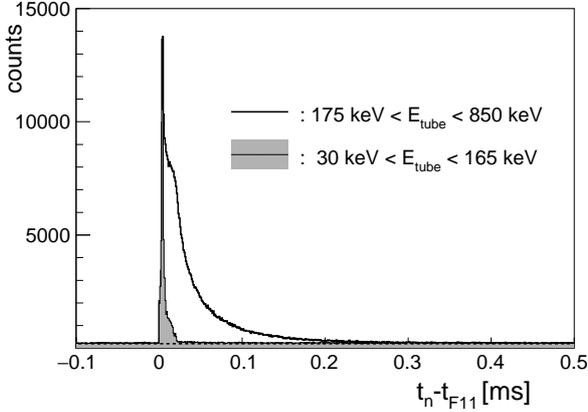}
 \end{center}
 \caption{Distribution of time differences between signals in BRIKEN $^{3}$He tubes identified
as valid neutron signals (see Fig.~\ref{fig:en3he}) and
F11 plastic signals ($t = t_{n} -t_{F11}$), represented as the unfilled histogram
(black line). The grey filled histogram
represents the time difference for $\gamma$-like signals in BRIKEN
$^{3}$He tubes selected in the energy interval [30~keV, 165~keV].}
 \label{fig:tf11n}
\end{figure}

\begin{figure}[ht]
 \begin{center}
 \includegraphics[width=7.8cm]{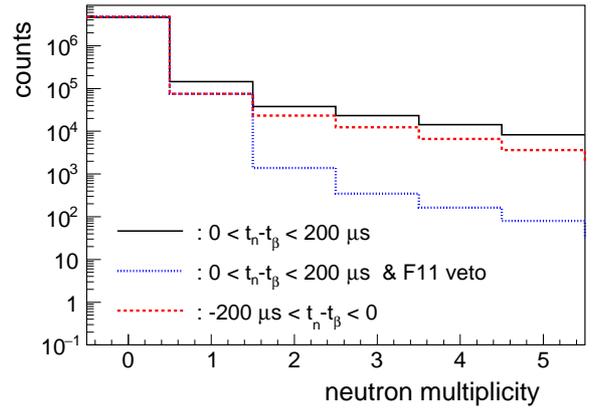}
 \end{center}
 \caption{Number of neutrons detected in BRIKEN within a 
time correlation window of $\Delta t_{\beta n} = 200 \mu$s
with respect to $\beta$ signals detected in AIDA. The black continuous line represent
the multiplicity distribution of
neutrons arriving after $\beta$ signals. The dotted blue line 
is the analogous distribution including the veto condition
that no signal was detected in the F11 plastic detector in the interval of
$\Delta t_{F11n} = 200$~$\mu$s before the neutron signal.  
The dashed red line represents the multiplicity of neutrons
arriving within $\Delta t_{\beta n} = 200 \mu$s before the $\beta$ signal, i.e. accidental 
coincidences with background neutrons.}
 \label{fig:nmult}
\end{figure}

We also observed that the beam-induced neutron background has a 
large multiplicity $M_{d}$ (number of tubes firing).
This can be a limitation for the measurement of  multiple neutron emission
probabilities. Figure \ref{fig:nmult} shows the observed multiplicity distribution
of neutrons coming within $\Delta t_{\beta n}$ after a $\beta$ signal during the
commissioning run (black continuous line).
It should be noted that in this run no multiple neutron
emitters were produced. The figure also shows the multiplicity histogram
obtained when the coincidence window is set \emph{before}
the $\beta$ signal, representing the accidentally correlated neutrons (red dashed line).
In this distribution the high multiplicity of background
neutrons is clearly seen: $M_{d}=2$ and 3  
are 30\%  and 16\% respectively of $M_{d}=1$.
When the F11 plastic veto condition is applied
a strong reduction of the higher multiplicities is obtained as observed
in Fig.~\ref{fig:nmult} (dotted blue line). The reduction factor is $\sim 2$ for $M_{d}=1$,
$\sim 30$ for $M_{d}=2$, and $\sim 70$ for $M_{d}=3$, 
demonstrating the usefulness of the veto.
A similar veto condition using the AIDA plastic detector 
can be added as well.
This will reduce the neutron background contribution 
associated with light particles in the beam 
that go through AIDA. Light particles remain undetected in the F11 plastic because
of the small thickness of this detector. For the $\mathrm{A} \sim 80$ run
during May-June 2017
the addition of the AIDA plastic veto condition yields a 20\%
further reduction of the background. In later experiments with heavier beams
the impact is larger. During the commissioning run the AIDA plastic veto condition
had no significant impact. Likewise we found no significant reduction
of the background vetoing with signals from the VETO plastic detectors attached
to the PE shielding.

In spite of the reduction, the large rate of background neutrons prevents us
from using direct neutron counting in the analysis. 
Decay neutron signals are buried in the background 
or result in a low accuracy. Therefore we have to rely on 
additional $\beta$-gating
as a way to improve the signal-to-background ratio.

\section{Analysis methodology for the extraction of $P_{xn}$ and $T_{1/2}$}
\label{sec:analysis}
The main goal of the analysis of BRIKEN data is to extract accurately the neutron emission
probability and half-life characterizing the decay of the implanted nuclei.
Actually both quantities come from the same analysis procedure,
although sometimes $T_{1/2}$ is already known from previous measurements 
with sufficient accuracy
and only $P_{xn}$ needs to be determined. This is a favorable situation 
because it reduces the uncertainty of the result.

To extract $P_{xn}$ we need to quantify, for a given implanted nucleus, the number of 
$\beta$ decays followed by the emission of $x$ neutrons and compare it 
with the total number of decays.
Since we do not know when an implanted ion is going to decay,
we can only associate decays with implants statistically
by constructing spatial and temporal correlations.
Thus for each identified implanted ion we construct the histogram $h_{i\beta} (t)$
of time differences $t=t_{\beta}-t_{ion}$
with \emph{all} $\beta$ events
occurring within the same spatial location
and within a specified time range. 
The truly correlated decays will stand out from a flat
background of uncorrelated decays.
To assess the probability of $\beta1n$ decays we need an additional histogram $h_{i\beta 1n} (t)$
similar to the previous one but adding the condition that
one neutron, and only one, was detected after the $\beta$ within the 
moderation-plus-capture time ($\Delta t_{\beta n} = 200 \mu$s).
For $\beta2n$ decays we introduce another histogram $h_{i\beta 2n} (t)$ with
the condition that two neutrons are detected within $\Delta t_{\beta n}$ after the $\beta$ particle.
And similarly for any other $\beta xn$ decay.

However, these histograms contain not only the counts from parent decays
but also from all descendants, in the case of $h_{i\beta} (t)$,
from descendants in the decay chain that are $\beta1n$ emitters 
in the case of $h_{i\beta 1n} (t)$, and so forth. Thus to disentangle
the parent and descendant contributions we must fit the histograms 
using the appropriate solution of the Bateman equations which describe
the time evolution of all activities. We use the generic form of the solution
proposed in \cite{skr74} which in our case simplifies to:

\begin{equation}
N_{k} (t) = N_{1}
 \prod_{i=1}^{k-1}  ( b_{i,i+1} \lambda_{i} ) \times \left( \sum_{i=1}^{k} 
 \frac{ e^{-\lambda_{i} t}}{\displaystyle \prod_{j=1 \neq i}^{k} (\lambda_{j}-\lambda_{i})}  
\right )
\label{eq:skrable}
\end{equation}

\begin{figure}[ht]
 \begin{center}
 \includegraphics[width=6.8cm]{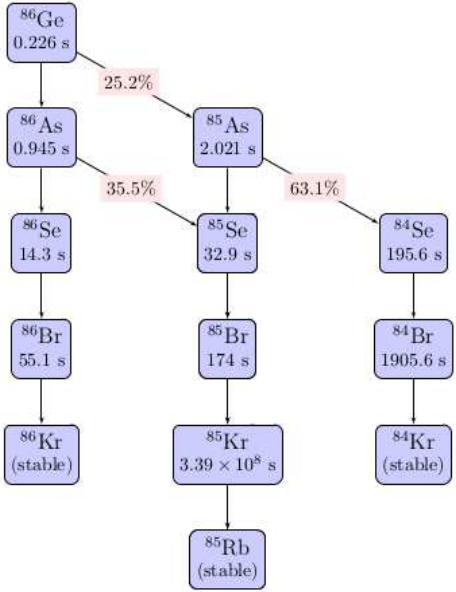}
 \end{center}
 \caption{Figure representing the various decay paths for the disintegration
of $^{86}$Ge. For each nucleus the half-life is indicated. 
The branchings indicate the $P_{1n}$ values.
Both numbers are taken from standard databases \cite{ensdf}.}
 \label{fig:network}
\end{figure}

$N_{k} (t)$ is the number of $k$-type nuclei in a given decay path
at time $t$, $N_{1}=N_{1}(t=0)$ is the initial 
number of implanted parent nuclei, and $\lambda=\ln 2/T_{1/2}$
is the decay constant.
The branching ratio  $b_{i,i+1}$ from nucleus $i$ to nucleus $i+1$ in the decay
chain defines the decay path.
In general these branchings are just the $P^{i}_{xn}$ with $x=0,1,2,...$~.
In the presence of isomers with sufficiently long half-life
that de-excite with a certain probability by internal transition (IT), 
the corresponding branching (decay path) must be included also.
A typical decay network with various branching points is represented in Fig.~\ref{fig:network}.

Obviously the fit function must also include the $\beta$ and neutron detection
efficiencies. As discussed in \cite{agr16} both efficiencies are energy dependent.
The $\beta$ efficiency depends on the end-point energy, 
$\varepsilon_{\beta} (Q_{\beta}-E_{x})$,
and the neutron efficiency depends on the neutron energy,
$\varepsilon_{n} (E_{n})$.
For the implant-$\beta$ time histogram $h_{i\beta} (t)$ the fit function takes
the form:

\begin{equation}
f_{i \beta} (t) =  \sum_{k \in \beta} \bar{\varepsilon}_{\beta}^{k} \lambda_{k} N_{k} (t)
\label{eq:fib}
\end{equation}

Here the summation runs over the parent and all decay descendants.
Since several decay paths can run through a given nucleus $k$
we must keep proper accounting of the inventory when computing
$N_{k} (t)$ using Eq.~\ref{eq:skrable}. The $\beta$ detection
efficiency for the $k$ nucleus is represented by
$\bar{\varepsilon}_{\beta}^{k}$. The bar symbol emphasizes that it is
obtained as a weighted average with the $\beta$-intensity distribution 
$I_{\beta} (E_{x})$ expressed as (dropping the index $k$ for clarity):

\begin{equation}
\bar{\varepsilon}_{\beta} =\frac{\int_{0}^{Q_{\beta}} I_{\beta} (E_{x}) 
\varepsilon_{\beta} (Q_{\beta}-E_{x}) dE_{x}}{\int_{0}^{Q_{\beta}} I_{\beta} (E_{x}) dE_{x}}
\label{eq:effb}
\end{equation}

Since $Q_{\beta}$ and $I_{\beta} (E_{x})$ vary from one nucleus 
to another, the average detection efficiency is nucleus dependent as indicated
in Eq.~\ref{eq:fib}.

For the $h_{i\beta 1n} (t)$ histogram the fit function takes the form:

\begin{equation}
f_{\beta 1n} (t) =  \sum_{k \in \beta 1n} \bar{\varepsilon}_{\beta 1n}^{k} 
\bar{\varepsilon}_{1n}^{k} P_{1n}^{k} \lambda_{k} N_{k} (t)
\label{eq:fib1n}
\end{equation}

Here the summation runs over the parent and all descendants that are $\beta 1n$
emitters. The $\beta$ efficiency in the $\beta 1n$ channel 
is averaged in the excitation energy range
$[S_{1n},Q_{\beta}] $ 
and is weighted by the $\beta$ intensity
leading to $1n$ emission $I_{\beta 1n} (E_{x})$, 
thus it is different from $\bar{\varepsilon}_{\beta}$ for the same nucleus:

\begin{equation}
\bar{\varepsilon}_{\beta 1n} =\frac{\int_{S_{1n}}^{Q_{\beta}} I_{\beta 1n} (E_{x}) 
\varepsilon_{\beta} (Q_{\beta}-E_{x}) dE_{x}}{\int_{S_{1n}}^{Q_{\beta}} I_{\beta 1n} (E_{x}) dE_{x}}
\label{eq:effb1n}
\end{equation}

The average neutron efficiency is weighted by the $\beta 1n$ 
neutron energy spectrum $I_{1n} (E_{n})$:

\begin{equation}
\bar{\varepsilon}_{1n} =\frac{\int_{0}^{Q_{\beta 1n}} I_{1n} (E_{n}) 
\varepsilon_{n} (E_{n}) dE_{n}}{\int_{0}^{Q_{\beta 1n}} I_{1n} (E_{n}) dE_{n}}
\label{eq:eff1n}
\end{equation}

For the $h_{i\beta 2n} (t)$ histogram the fit function takes the form:

\begin{equation}
f_{\beta 2n} (t) =  \sum_{k \in \beta 2n} \bar{\varepsilon}_{\beta 2n}^{k} 
(\bar{\varepsilon}_{2n}^{k})^{2} P_{2n}^{k} \lambda_{k} N_{k} (t)
\label{eq:fib2n}
\end{equation}

The summation runs over the parent and all descendants that are $\beta 2n$
emitters, and the average $\beta$ and neutron detection efficiencies 
for the $\beta 2n$ channel take the form:

\begin{equation}
\bar{\varepsilon}_{\beta 2n} =\frac{\int_{S_{2n}}^{Q_{\beta}} I_{\beta 2n} (E_{x}) 
\varepsilon_{\beta} (Q_{\beta}-E_{x}) dE_{x}}{\int_{S_{2n}}^{Q_{\beta}} I_{\beta 2n} (E_{x}) dE_{x}}
\label{eq:effb2n}
\end{equation}

\begin{equation}
\bar{\varepsilon}_{2n} =\frac{\int_{0}^{Q_{\beta 2n}} I_{2n} (E_{n}) 
\varepsilon_{n} (E_{n}) dE_{n}}{\int_{0}^{Q_{\beta 2n}} I_{2n} (E_{n}) dE_{n}}
\label{eq:eff2n}
\end{equation}

Note that $\bar{\varepsilon}_{2n}$ is the efficiency for detection of one neutron
from the $\beta 2n$ channel.
For simplicity of notation in Eq.~\ref{eq:fib2n} we assume that $\bar{\varepsilon}_{2n}$ 
is the same for the two neutrons emitted, i.e. they have the same neutron intensity distribution.

From the definitions above it is clear that 
$\bar{\varepsilon}_{\beta 2n} \neq \bar{\varepsilon}_{\beta 1n}  \neq \bar{\varepsilon}_{\beta}$
and that $\bar{\varepsilon}_{2n} \neq \bar{\varepsilon}_{1n}$.
The formulae can be extended easily to $\beta 3n$, $\beta 4n$, ... decays.

The fact that all average $\beta$ and neutron efficiencies
are in principle different represents a challenge when extracting 
$P_{xn}$ and $T_{1/2}$ from the fit. 
There is no clear way
to determine these efficiencies for most of the decays, since the $\beta$ intensity
distributions and neutron energy spectra are not known. 
In some cases there are general arguments, related to the size of the decay windows
and expected shape of the intensity distributions, 
that allow us to assume that all $\beta$ efficiencies 
are equal and/or all neutron efficiencies are equal. 
In this situation  $\bar{\varepsilon}_{\beta}$ factors out and only $\bar{\varepsilon}_{n}$
is needed to perform the fit.
However this assumption
can introduce systematic errors that need to be studied and quantified.
Examples of this will be presented later.

One can see from the form of Eq. \ref{eq:fib}, \ref{eq:fib1n}, and \ref{eq:fib2n}, 
that the parent decay half-life intervenes in the shape of the three histograms
$h_{i\beta} (t)$, $h_{i\beta 1n} (t)$ and $h_{i\beta 2n} (t)$.
The same is true for $P_{1n}$ and $P_{2n}$ which appear explicitly
in the last two equations, but implicitly in all three through  the parent
decay branchings ($b_{1,2}$, see Eq.~\ref{eq:skrable}) 
that determine the weight of the respective
descendant decays. The best way to take into account these correlations
is to perform a simultaneous fit to all three histograms, 
where the unknown $P_{1n}$, $P_{2n}$, ... and $T_{1/2}$ are the parameters of the fit. 
An additional fit parameter
representing the normalization is always needed. 
This is $N_{1}$, the initial number of implanted parent nuclei.
However, before the fit can be performed we must take 
into account various background contributions 
to the experimental histograms, as explained in the next Section.

\section{Background correction}
\label{sec:background}
A number of background sources affect the experimental histograms
$h_{i\beta} (t)$ and $h_{i\beta xn} (t)$. Signals identified as $\beta$
signals in AIDA which are not related to the decay of the implanted
nucleus contribute to the accidental $\beta$ background. 
It affects all histograms and has a flat time distribution.
This uncorrelated $\beta$ background comes from: 1) $\beta$ particles
belonging to the decay chain of other nuclei implanted in the
same correlation area, 2) light particles that pass through the detector and leave an 
energy similar to $\beta$ particles, 3) detector noise.
This background component imposes a limit on the minimum
detectable activity and can be reduced by optimizing
the implant-$\beta$ correlation area, vetoing the signals 
correlated with the AIDA plastic, and reducing noise and optimizing
thresholds in AIDA.

\begin{figure}[ht]
 \begin{center}
 \includegraphics[width=7.8cm]{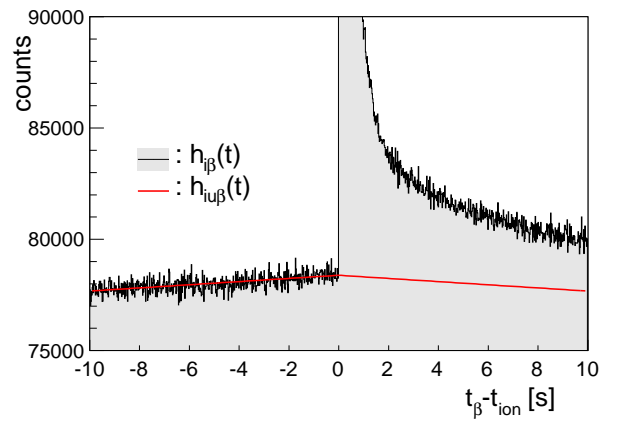}
 \end{center}
 \caption{Zoom on the implant-$\beta$ time correlation histogram
$h_{i\beta} (t)$, $t = t_{\beta} -t_{ion}$, for $^{83}$Ga
in the time range from -10~s to +10~s. The background is fitted
with a linear function (red line) for $t<0$ and symmetrically extrapolated to $t>0$.}
 \label{fig:buncorr}
\end{figure}

Our way to determine this background component
is to: 1) construct backwards in time implant-$\beta$ correlations ($t<0$),
where only the uncorrelated $\beta$ particles contribute, 
and 2) extrapolate to positive times.
An example of this is shown in Fig.~\ref{fig:buncorr}
showing the $h_{i\beta} (t)$ histogram for $^{83}$Ga in the time
range $[-10~s, +10~s]$.
As can be seen, the background time distribution is not constant for $t<0$
and has a small positive slope. This effect could be traced
back to accidental beam interruptions
during a run, when the $\beta$ rate decreases. 
During the commissioning experiment there were frequent
beam interruptions (instabilities) lasting from less than a second
to few tens of seconds.
We verified that removing from the time correlation 
the data coming up to 10~s before and after 
a beam interruption the uncorrelated background becomes nearly constant
reducing the statistics by a factor of 2.
Using MC simulations we verified that the background shape
depends on half-lives and length of the interruption.
For a random distribution of interruption intervals we obtained 
a background shape that is symmetrical around t=0  to a good
approximation, thus we take this assumption in our analysis
(see Fig. \ref{fig:buncorr}).
It is worth to mention that the analysis of the data obtained removing beam interruptions
gives the same result within statistics than the full data set (see Section \ref{sec:results}).
We observe that in most of the cases a linear function provides a good reproduction
of the uncorrelated $\beta$ background. On occasions an exponential function 
reproduces the shape better. In either case they define the
correction histograms $h_{i u \beta} (t)$ and $h_{i u \beta xn} (t)$ .

In the case of the $h_{i\beta} (t)$ histogram this is the only background
contribution thus the relation of the measured histogram 
to the unperturbed time distribution $f_{i \beta} (t)$ (see Eq.~\ref{eq:fib}) is given by:

\begin{equation}
h_{i\beta} (t)= f_{i\beta} (t) + h_{i u \beta} (t)
\label{eq:c1b}
\end{equation}

Another important source of background comes from neutrons
that are accidentally correlated with $\beta$ particles within
the time window for $\beta$-neutron correlation $\Delta t_{\beta n}$.
This only affects the $h_{i\beta xn} (t)$ histograms.
Neutron signals that can be accidentally correlated come from:
1) neutrons emitted by other implanted nuclei,
2) beam induced neutrons, 3) ambient background neutrons, and
4) detector noise, including $\gamma$-like signals in $^{3}$He tubes.
This background  component affects the minimum detectable $\beta xn$
activity and can be minimized with proper detector shielding,
discrimination of beam induced neutron signals and detector noise reduction.
A characteristic of this type of background is that it follows
the time distribution of implant-$\beta$ correlations and thus it has a time
structure. This has a direct impact on the extraction of  
$P_{xn}$ and $T_{1/2}$ from the fit and it is crucial to have
an accurate method of background correction.

\begin{figure}[ht]
 \begin{center}
 \includegraphics[width=7.8cm]{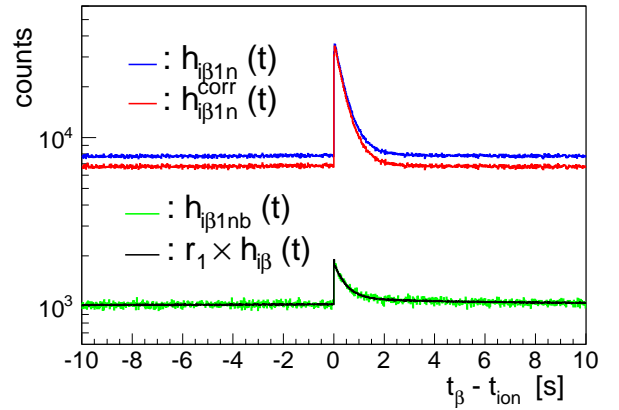}
 \end{center}
 \caption{Different implant-$\beta$ time correlation histograms for $^{83}$Ga.
Blue: uncorrected implant-$\beta$-1n time distribution $h_{i\beta 1n} (t)$; 
green: implant-$\beta$ time distribution of $\beta$ particles in accidental coincidence with 
one background neutron $h_{i\beta 1nb} (t)$;  
black: scaled implant-$\beta$ time distribution $r_{1} h_{i\beta} (t)$; 
red: corrected implant-$\beta$-1n time distribution $h_{i\beta 1n}^{corr} (t)$. See text for details.}
 \label{fig:b1ncorr}
\end{figure}

\begin{figure}[ht]
 \begin{center}
 \includegraphics[width=7.8cm]{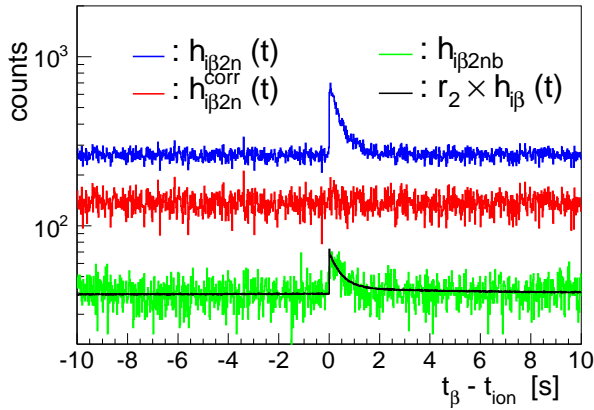}
 \end{center}
 \caption{Different implant-$\beta$ time correlation histograms for $^{83}$Ga.
Blue: uncorrected implant-$\beta$-2n time distribution $h_{i\beta 2n} (t)$; 
green: implant-$\beta$ time distribution of $\beta$ particles in accidental coincidence with 
two background neutrons $h_{i\beta 2nb} (t)$;  
black: scaled implant-$\beta$ time distribution $r_{2} h_{i\beta} (t)$; 
red: corrected implant-$\beta$-2n time distribution $h_{i\beta 2n}^{corr} (t)$. See text for details.}
 \label{fig:b2ncorr}
\end{figure}

We introduce here a new method of correction for accidental
$\beta$-neutron background that estimates accurately
its contribution directly from the data. 
Figures \ref{fig:b1ncorr} and \ref{fig:b2ncorr} show the relevant 
histograms for the discussion
in the example of $^{83}$Ga decay.
In this case we use the data taken during the May-June 2017 run,
which has much higher statistics ($5 \times 10^{6}$ implanted ions)
to demonstrate the results.

The method is based
on the use of backwards in time $\beta$-neutron correlations
to determine the number of accidental neutrons correlated with $\beta$ particles.
The idea is that the number of accidental neutrons
coming within  $\Delta t_{\beta n}$ after the $\beta$ is 
on average the same as the number of neutrons coming within  
$\Delta t_{\beta n}$ \emph{before} the $\beta$, all of which
are of necessity accidentals. The only assumption
here is that the neutron background rate is not changing,
on average, over a period of a few hundreds of $\mu$s.
We construct a new implant-$\beta$ 
time correlation histogram with the condition that one neutron arrives
within $\Delta t_{\beta n}$ before the $\beta$.
This histogram, that we designate $h_{i\beta1nb} (t)$
is shown in Fig.~\ref{fig:b1ncorr}
in green. Notice that the shape of this histogram is identical to
the scaled $h_{i\beta} (t)$ histogram represented in black in the figure.
The scaling factor $r_{1}$:

\begin{equation}
 r_{1} = \frac{\int_{-10 \mathrm{s}}^{+10 \mathrm{s}} h_{i\beta1nb} (t) dt}
{\int_{-10 \mathrm{s}}^{+10 \mathrm{s}} h_{i\beta} (t) dt}
\label{eq:r1}
\end{equation}

is the probability of having one-accidental-neutron per detected $\beta$, determined 
with great precision because we use the full statistics of the histograms.
In the present example $r_{1} = 0.013173 (13)$.
The value of $r_{1}$ changes by a few percent from one nucleus
to another due to changes in the relative background conditions.
For example a nucleus with high implantation rate and large $P_{n}$ sees
less background than a nucleus with low implantation rate and small $P_{n}$.

We also construct $h_{i\beta2nb} (t)$, the implant-$\beta$ time correlation histogram
with the condition that two neutrons are coming within
$\Delta t_{\beta n}$ before the $\beta$. 
This is shown in green in Fig.~\ref{fig:b2ncorr} and as before
its shape is matched by the scaled  $h_{i\beta} (t)$ histogram (in black).
The scaling factor $r_{2}$:

\begin{equation}
 r_{2} = \frac{\int_{-10 \mathrm{s}}^{+10 \mathrm{s}} h_{i\beta2nb} (t) dt}
{\int_{-10 \mathrm{s}}^{+10 \mathrm{s}} h_{i\beta} (t) dt}
\label{eq:r2}
\end{equation}

represents the probability of having two-accidental-neutrons per detected $\beta$.
In the present example $r_{2} = 0.0005056 (25)$,
twenty five times smaller than $r_{1}$. A similar procedure can be applied
for higher accidental neutron multiplicities. The
red-dashed histogram in Fig.~\ref{fig:nmult}, representing the 
multiplicity of neutrons accidentally correlated with a $\beta$ particle,
give us information about the value of $r_{n}$ for $n>2$.
The total probability of accidental neutrons per detected $\beta$ is
$r = r_{1} + r_{2} + ...$ .

Let us consider the case of decays followed by one-neutron emission.
The measured histogram $h_{i\beta1n} (t)$ is represented
in blue in Fig.~\ref{fig:b1ncorr}. This histogram
has to be corrected for background contributions to obtain
the unperturbed time distribution represented by the function $f_{i\beta 1n} (t)$
defined in Eq.~\ref{eq:fib1n}.
Accidental neutron coincidences have two effects on this distribution. 
One effect is a loss of counts whenever one or more background neutrons
comes accidentally within $\Delta t_{\beta n}$ after the $\beta$
in addition to the truly correlated neutron.
The loss is proportional
to $r$ the total probability of accidental neutrons per detected $\beta$.
The net effect is a scaling down of the distribution, of the form $(1-r) f_{i\beta 1n} (t)$. 
The other effect is the appearance of spurious counts in the histogram
when one accidental neutron
correlates with  $\beta$ particles that do not see correlations with decay neutrons.
The latter have a time distribution that can be obtained 
as the difference between the distribution of all
detected $\beta$ events, $f_{i\beta} (t)$, and the distribution of $\beta$ events
where one decay neutron was detected, $f_{i\beta 1n} (t)$. Scaling this distribution
by $r_{1}$, the probability of one-accidental-neutron per $\beta$,
gives the contribution $r_{1} (f_{i\beta} (t)-f_{i\beta 1n} (t))$.
The measured histogram is then the sum of both terms plus the uncorrelated
background contribution $h_{i u \beta 1n} (t)$. After some rearrangement it gives:

\begin{equation}
h_{i\beta 1n} (t)= (1-r-r_{1}) f_{i\beta 1n} (t) + r_{1} f_{i\beta} (t) + h_{i u \beta 1n} (t)
\label{eq:c1b1n}
\end{equation}

To visualize the size of the corrections it is useful to calculate the histogram 
$h_{i\beta1n}^{corr} (t)=(h_{i\beta1n} (t)-r_{1}h_{i\beta} (t)) / (1-r-r_{1})$
that is shown in red in Fig.~\ref{fig:b1ncorr}.
Note that both $h_{i\beta1n} (t)$ and $h_{i\beta} (t)$ include their
respective uncorrelated backgrounds. As can be seen in Fig.~\ref{fig:b1ncorr}
the correction is small in this case but it would be important if the $P_{1n}$
value is small. An example will be shown later.

Let us turn now to the case of two-neutron emission.
The measured $h_{i\beta 2n} (t)$ histogram 
is represented in blue in Fig.~\ref{fig:b2ncorr}.
This histogram has to be corrected for background contributions to obtain
the unperturbed time distribution represented by the function $f_{i\beta 2n} (t)$
defined in Eq.~\ref{eq:fib2n}. The effect of accidental coincidences with
background neutrons in this histogram
is similar to the one explained above: loss of true counts
and appearance of spurious counts. In addition one has to 
modify the corrections to the $h_{i\beta 1n} (t)$ histogram  (Eq.~\ref{eq:c1b1n})
to take into account the contribution of the $\beta$2n decay channel \cite{azu80}. 
In \ref{sec:appendix} we explain in detail how to obtain the different correction terms. 
Here we simply give the result expressed as the relation
between the measured histograms $h_{i\beta 1n} (t)$ and $h_{i\beta 2n} (t)$ 
and the unperturbed time distributions $f_{i\beta} (t)$, $ f_{i\beta 1n} (t)$ and 
$f_{i\beta 2n} (t)$:

\begin{equation}
\begin{split}
h_{i\beta 1n} (t) = & (1-r-r_{1}) f_{i\beta 1n} (t) + r_{1} f_{i\beta} (t)\\
&  + (2 r_{e} (1-r-r_{1})-r_{1}) f_{i \beta 2n} (t) + h_{i u \beta 1n} (t)
\label{eq:c2b1n}
\end{split}
\end{equation}

\begin{equation}
\begin{split}
h_{i\beta 2n} (t) = & (1-r-r_{2} + 2 r_{e} (r_{1}-r_{2})) f_{i\beta 2n} (t) \\
 & + (r_{1} -r_{2}) f_{i \beta 1n} (t) + r_{2} f_{i\beta} (t)  + h_{i u \beta 2n} (t)
\label{eq:c2b2n}
\end{split}
\end{equation}

where $r_{e} = (1-\bar{\varepsilon}_{2n}) / \bar{\varepsilon}_{2n}$.

The computation of the background corrected implant-$\beta$-1n and  implant-$\beta$-2n
histograms from the measured histograms gets more complicated now 
because of the interdependence of the corrections.
We give in \ref{sec:appendix} the appropriate formulas.
In the present example, the corrected histogram $h^{corr}_{i\beta 2n} (t)$ is represented
in red in Fig.~\ref{fig:b2ncorr}.
As can be observed the peak in the uncorrected time distribution (blue)
disappears in the corrected time distribution, which is completely flat.
This agrees with the fact that $^{83}$Ga must have a extremely small $P_{2n}$ 
due to the small $Q_{\beta 2n}=0.89$~MeV. 
It confirms the accuracy of the correction method
and demonstrates the importance of accidental
neutron background correction for determining small $P_{2n}$ values.

Similar formulas for $\beta 3n$ emitters are given in \ref{sec:appendix}.

We should mention that
an alternative method of analysis and background correction for
BRIKEN data has been developed \cite{ras18}. Compared to the method
presented here, this alternative method determines initial parent activities
for each of the $xn$ decay channels
from independent fits to the corresponding time correlation histograms. 
These initial activities are then combined to obtain $P_{xn}$ values
applying global time-independent corrections for the correlated neutron
background contribution.

\section{Selected results}
\label{sec:results}

We present in this section details of the analysis for a few isotopes in order
to illustrate the procedure and the quality of results.

The data was acquired during the commissioning run over
10 effective hours of measurement at a primary beam intensity of 20~pnA.
In the sort of AIDA data, $\beta$ events are treated by defining clusters of 
consecutive strips firing above the noise threshold
(strip dependent) in both X and Y directions. 
This takes into account the fact that $\beta$ particles can have a long
range in Si.
A $\beta$ pixel is determined by the energy weighted centroid of the cluster of strips
with the condition that X and Y energies are similar. Implantation events
have a small strip multiplicity (one or two strips) and are defined by the last layer
(DSSD) firing the low gain electronic branch. We consider only ion-$\beta$ correlation events 
when they happen in the same layer (Z position) and the difference of  X and Y 
centroid positions between $\beta$ and ion is less than three strips (defining a correlation area of 
$3.3~\mathrm{mm} \times 3.3~\mathrm{mm}$).

We discovered during the run in May-June 2017 a problem related to the 
design of the AIDA adaptor PCB cards
that serve to connect the flat cables coming from the Si DSSD . 
The effect was a transient induced by implantation events in the high gain
electronics  which is interpreted as a $\beta$ event. 
The effect lasted up to a few tens of ms and  appears as a spurious implant-$\beta$
time correlation extending up to 30-40~ms. These background signals 
can be effectively eliminated by neglecting the first 50~ms in the fit of the
time correlated histograms. In the $\mathrm{A} \sim 80$ runs this is not an issue
because the half-lives are relatively long. After identification of the problem
the coupling cards were modified and the effect eliminated as verified
during the October-November 2017 run.

\subsection{Fitting procedure}
\label{sec:fitting}
 
We construct the time correlation histograms 
$h_{i\beta} (t)$ and $h_{i\beta 1n} (t)$ for all implanted ions that are identified. 
We choose a time window from -10~s to +10~s
that is appropriate for all the cases analyzed. The binning of the histograms 
for each nucleus is chosen 
balancing the need to have enough points to determine the activity evolution and
minimize the statistical fluctuation in the bin counts. 

A fitting subroutine was written using ROOT::Fit classes \cite{root}.
The inputs to the program are the measured time correlation histograms,
the half-life and neutron emission probabilities
of all nuclei involved and
the corresponding $\beta$ and neutron efficiencies.
All parameters have an associated uncertainty
and can be fixed during the fit.
The program automatically reconstructs the decay network based on the 
nuclei and $P_{xn}$ information provided.

For the fit we do not subtract the different
background contributions from the measured histograms but rather
include these contributions in the fit function. See \ref{sec:appendix}.
This is the proper way to handle the corrections in view of the use
of Maximum Likelihood estimators.
Histogram subtraction destroys the Poisson character of bin counts
leading eventually  to negative counts for low statistics. 
In general we use the Binned Maximum Likelihood (BML) algorithm to fit the 
histograms, except when the 
very low statistics suggest the use of the Unbinned Maximum Likelihood (UML).
In this case the event data are provided in list mode. 
The uncorrelated $\beta$-ion background 
is obtained from a fit to the negative time range
for each of the $h_{i\beta} (t)$ and $h_{i\beta xn} (t)$ histograms
taking into account the effect of the correlated neutron background correction histograms
(see \ref{sec:appendix}). 
The fit to the positive time range skips
the first few bins in order to exclude the initial 50~ms range where the ion induced $\beta$
background appears. 

To evaluate the systematic
uncertainty due to the parameters fixed during the fit 
(half-lives, neutron branchings, backgrounds, efficiencies) we use a Monte Carlo
approach.
For any chosen subset of parameters we define a multivariate
normal distribution, using the adopted value of each parameter as the mean and 
the square of the quoted uncertainty as the variance. In general we assume that different
parameters are uncorrelated (diagonal covariance matrix). 
The multivariate distribution is randomly sampled and the fit performed.
The resulting fit parameters ($P_{1n}$, $T_{1/2}$, ...) are histogrammed  
and at the end the standard deviation of the sample distribution (eventually asymmetric)
is evaluated and quoted as the systematic uncertainty. 

For the fit we need to define $\beta$ and neutron efficiencies.
We assume that the nominal value  of the neutron efficiency is 
$\bar{\varepsilon}_{n} =66.8(20)$\% as discussed in Section \ref{sec:setup}.
This efficiency has to be renormalized in order to take into account the neutron count loss
because of the finite size of the $\beta$-neutron correlation window (0.43\%),  
and the dead-time introduced in the analysis 
by the neutron veto from the F11 plastic detector (8.79\%). This gives a value of 60.7(18)\%
for the effective neutron efficiency. This efficiency would have to be modified
for decays with a particular hard neutron spectrum.
The influence of $\beta$ efficiencies will be discussed in the next subsection.

\subsection{Effect of $\beta$ efficiencies}
\label{sec:beff}
\begin{figure}[ht]
 \begin{center}
 \includegraphics[width=7.8cm]{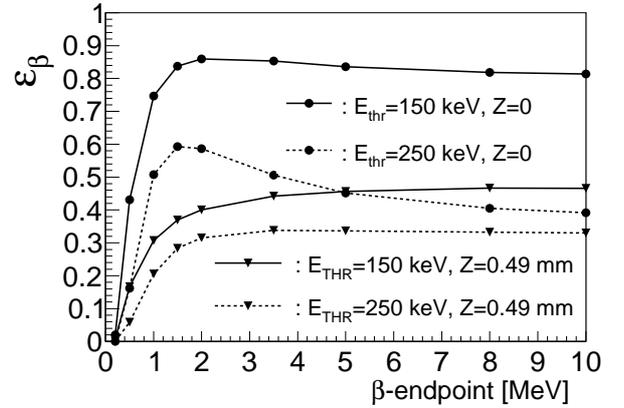}
 \end{center}
 \caption{Simulated $\beta$ efficiency in one of AIDA DSSDs as a function of $\beta$ 
endpoint energy. 
Circles: implantation at the center. Triangles: implantation close to the surface. 
Continuous line: Single strip lower energy threshold of 150~keV. 
Dashed line: Single strip lower energy threshold of 250~keV.}
 \label{fig:beff}
\end{figure}

The continuum nature of the $\beta$ spectrum together with the unavoidable 
minimum electronic thresholds introduce a $\beta$ end-point  energy dependence
in the $\beta$ detection efficiency \cite{agr16}. 
In the case of an implantation-decay
detector like AIDA the energy dependence is further complicated 
with a dependence on the implantation depth and with the 
method of reconstructing  $\beta$ events. 
To illustrate the dependence with threshold and implantation depth 
we show in Fig.~\ref{fig:beff} the result of Geant4 simulations
using the AIDA Si DSSD geometry. 
This efficiency does not include the effect of event
reconstruction and the absolute values are not representative of 
the actual $\beta$ efficiencies.

As can be observed in Fig.~\ref{fig:beff}
there is a fast drop in the efficiency for end-point energies below $1-2$~MeV.
When the $\beta$ particle is emitted from
the middle of the DSSD the efficiency is quite large if the threshold is low 
(below about 150~keV). In this case increasing the threshold (250~keV in the example) 
has a substantial effect, with the efficiency dropping for increasing end-point energies. 
When the implantation occurs close to one of the DSSD surfaces, half of the
$\beta$ particles have little chance of depositing enough energy and the
efficiency drops. The effect of a threshold increase is smaller 
in this case.
Because of the energy spread of implanted ions the implantation depth effect
is partially smeared out.
As explained in Section \ref{sec:analysis} this energy dependence can introduce differences 
in average efficiencies $\bar{\varepsilon}_{\beta}$ 
for different nuclei and decay branches, depending on the
$\beta$ intensity distribution,  which leads to
systematic errors in the results of the fit. Note that 
the systematic effect is due to the relative differences
and not to the absolute efficiency values.

It is not easy to determine experimentally the efficiency for every decay mode 
contributing significantly to the fit. One possibility is to use
the intensity of decay $\gamma$-rays observed in the CLOVER detectors 
to obtain information on the average $\beta$ efficiency.
This requires the comparison of $\beta$-gated with 
ungated $\gamma$ ray spectra \cite{tes16}, but it is in practice difficult to apply
because of the large background and the limited statistics.
Another approach is to calculate a realistic $\beta$ efficiency curve
from Monte Carlo simulated data and use $\beta$ intensity distributions
to compute the average efficiencies (Section \ref{sec:analysis}).
Since for most of the exotic decays this information is unknown
or poorly known, one must rely on theoretical $\beta$-strength distributions
to obtain an estimate. In spite of the uncertainties inherent in this approach
it can give a representative value of the size of the systematic error.
A third approach is to determine the $\beta$ efficiencies from the
time correlation data as will be discussed below.

\begin{figure}[ht]
 \begin{center}
 \includegraphics[width=7.8cm]{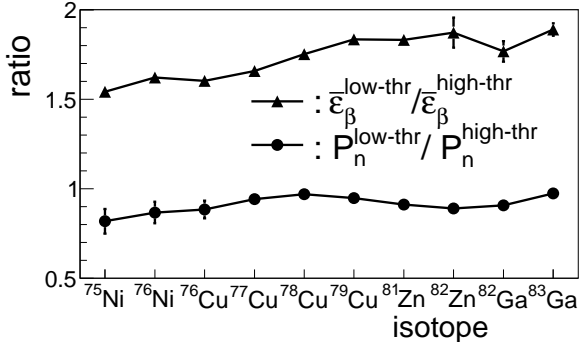}
 \end{center}
 \caption{Ratio of $\beta$ efficiencies (triangles) and $P_{1n}$ values (circles) 
obtained from the analysis of data sorted with two different $\beta$ energy
thresholds in AIDA (a low and a high threshold).}
 \label{fig:effpn}
\end{figure}

An important consideration is that the end-point energy dependence of
 $\varepsilon_{\beta}$ decreases as the threshold decreases
(it disappears at threshold zero). Therefore minimizing the 
effective $\beta$-energy threshold in AIDA data is
an important requirement to minimize this kind of systematic error.
One can test the magnitude of the systematic error by analyzing data
obtained with different $\beta$ thresholds.
Such a test is shown in Fig.~\ref{fig:effpn} for a set of Ni, Cu, Zn and Ga
 isotopes measured  during the commissioning run. They span
ranges of $Q_{\beta} = 9.4 - 13$~MeV and $Q_{\beta 1n} = 3.9 - 8.1$~MeV.

The fit to the time correlation histograms $h_{i\beta} (t)$ and $h_{i\beta 1n} (t)$
used to extract the $P_{1n}$ and $\beta$ efficiencies
shown in Fig.~\ref{fig:effpn},
assumes that all $\bar{\varepsilon}_{\beta}$ are equal. 
In this case the $\beta$ efficiency 
factors out of the fit function (see Eq.~\ref{eq:fib} and Eq.~\ref{eq:fib1n})
and can be determined  from the result of the fit.
The fit parameters are $P_{1n}$ and the normalization constant, equal to 
$\bar{\varepsilon}_{\beta} N_{1}$.
The remaining parameters are kept fixed to the adopted values in standard databases \cite{ensdf}.
Dividing the normalization constant by $N_{1}$ 
determined as the number of identified implanted ions
we obtain $\bar{\varepsilon}_{\beta}$.
The extracted efficiency is indicative of the effective $\beta$ threshold.
In the sort with the low threshold the $\beta$ efficiencies vary between 30\% and 42\%.
In the high threshold sort the efficiency range is 20-26\%.
The actual thresholds applied to AIDA $\beta$ data are strip dependent 
and vary between 100~keV and 250~keV in the high threshold sort
and between 50~keV and 200~keV in the low threshold sort.

Figure \ref{fig:effpn} shows the ratio of $\beta$ efficiencies and the ratio of $P_{1n}$ values
between the low threshold sort and the high threshold sort. The low threshold
increases $\bar{\varepsilon}_{\beta}$ between 54\% and 89\% with respect to the
high threshold. The impact on $P_{1n}$ is a reduction of all values
by factors from 2.5\% to 18\%.
We have not observed a clear correlation of the reduction factor 
with the size of the decay windows $Q_{\beta}$ and $Q_{\beta 1n}$.

Minimizing the thresholds in the AIDA sort is a challenging task because of the 
large number of channels and the nature of the noise, which is channel specific
and time dependent. A compromise must be established between lowering
the threshold and keeping a reasonable signal-to-noise ratio. 
For the commissioning run we adopt the low threshold sort discussed above 
that should reduce the effect of $\beta$ efficiency dependence in the data.
The question is whether a residual effect still remains.

\begin{figure}[ht]
 \begin{center}
 \includegraphics[width=7.8cm]{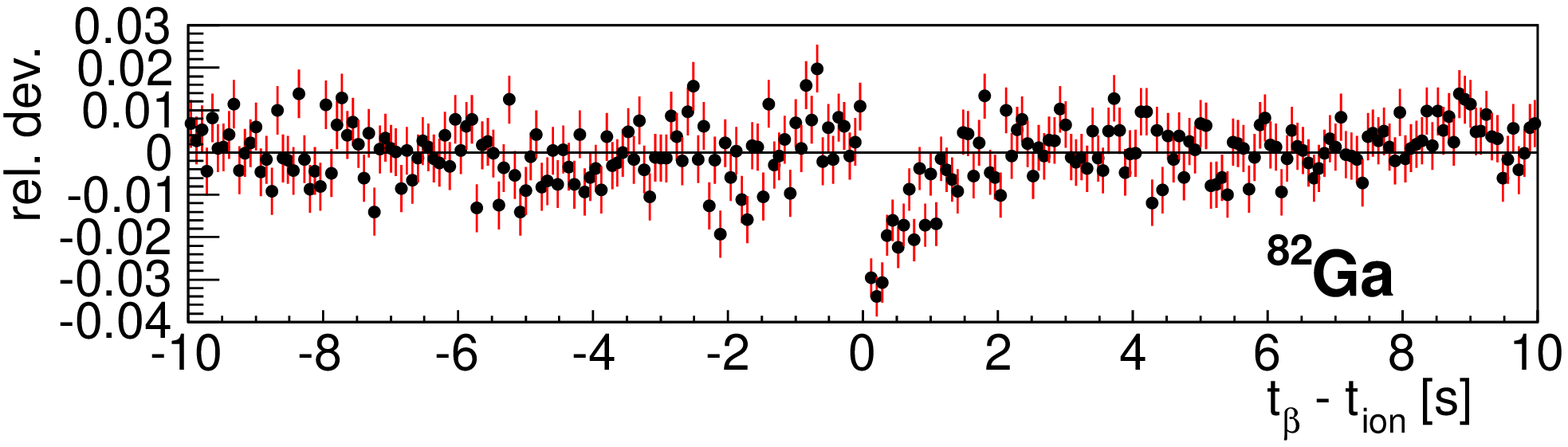}
 \includegraphics[width=7.8cm]{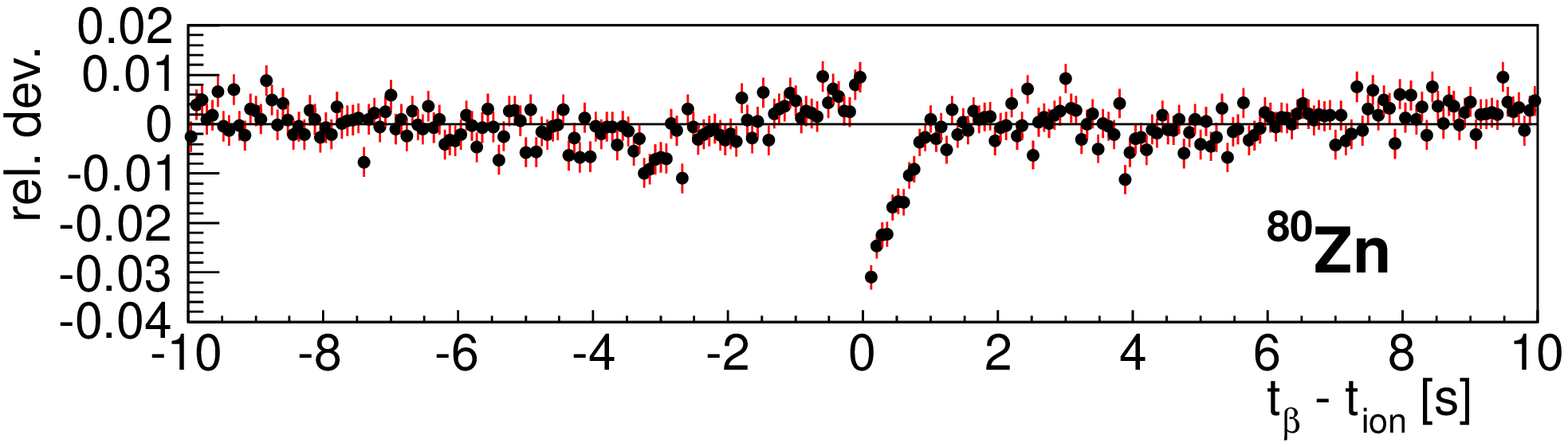}
 \includegraphics[width=7.8cm]{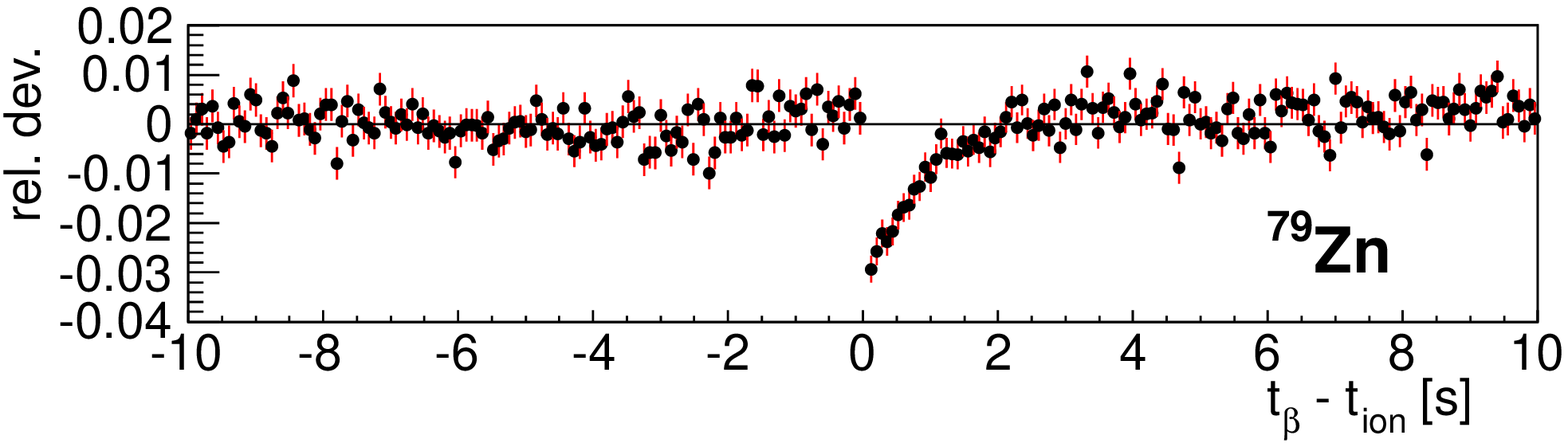}
 \includegraphics[width=7.8cm]{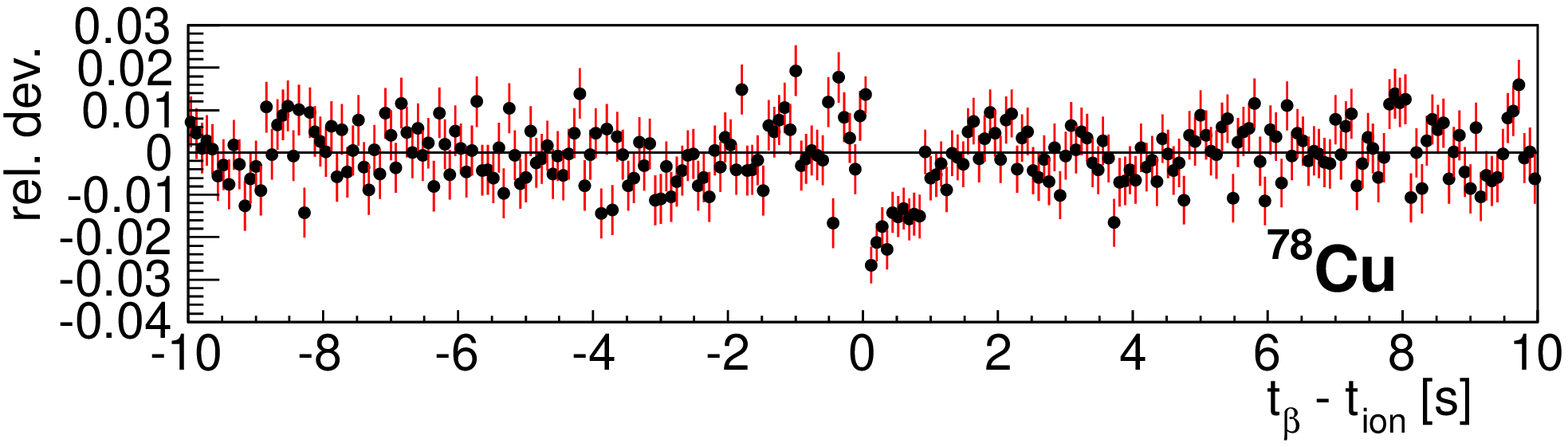}
 \includegraphics[width=7.8cm]{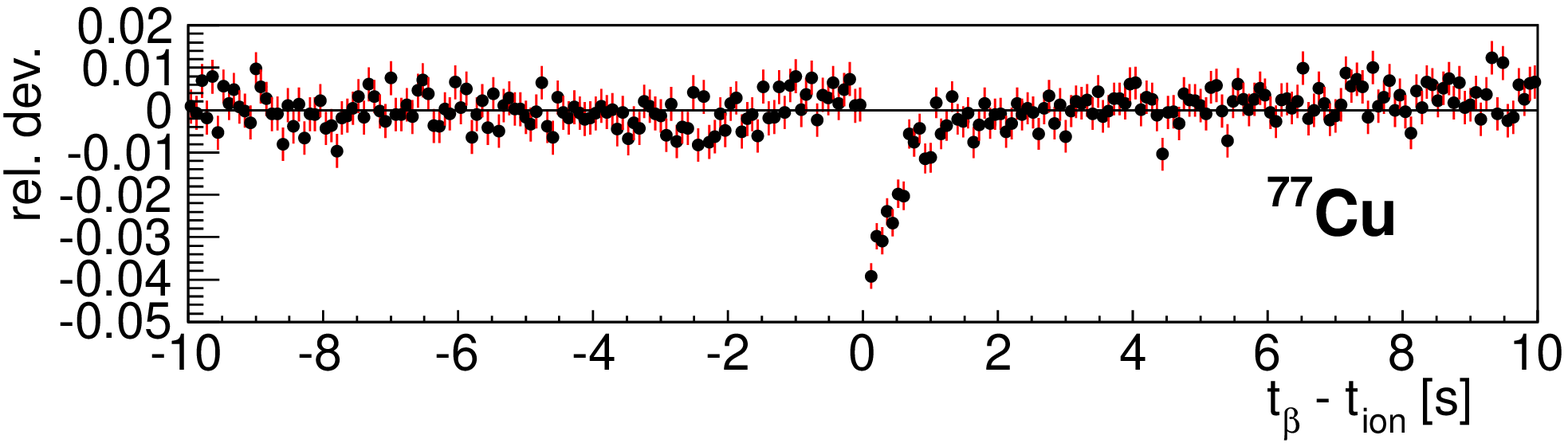}
 \end{center}
 \caption{Relative deviation between implant-$\beta$ time correlation data and the
fit function for isotopes with large implantation statistics during the commissioning ru.}
 \label{fig:fitdev}
\end{figure}

As a matter of fact we observe a small but systematic deviation between data and best fits 
for nuclei with high implantation statistics during the commissioning run.
Figure \ref{fig:fitdev} shows, relative values of fit residuals for 
implant-$\beta$ time correlation histograms
$(h_{i \beta} (t) - f_{i \beta} (t))/f_{i \beta} (t)$.
To show the effect more clearly the fit region is restricted to implant-$\beta$ correlation
times in the range [1~s, 10~s]. 
As can be observed all show a similar pattern: there is a deficit of
counts at short correlation times and a slight excess at long correlation times.
We do not observe this effect in the fit of the neutron-gated
implant-$\beta$ time correlation histogram $h_{i \beta 1n} (t)$.
We interpret this result as a consequence of the difference
in $\beta$ efficiency between the parent nucleus and descendants. In all cases except
$^{80}$Zn, this can be related to the much larger decay window $Q_{\beta}$ of the
parent.  The case of $^{80}$Zn will be commented on later.

In view of this result we will include in the fit, when necessary, as an additional adjustable
parameter the relative $\beta$ efficiency of selected decay modes in the chain.

\subsection{$^{82}$Ga: verification of neutron efficiency}
\label{sec:82ga}

During the commissioning run we accumulated $3.5 \times 10^{5}$ ions of $^{82}$Ga,
a good case to verify the neutron efficiency since the $P_{1n}$ for this decay
is fairly well known. There are three previous measurements which give consistent values:
21.4(22)\% \cite{lun80}, 19.8(10)\% \cite{war86}, and 22.2(20)\% \cite{tes16}.
Their weighted average gives 20.4(12)\%.
In addition there are two values with larger uncertainty that deviate significantly
from the other results, 31.1(44)\% from \cite{rud93} and  30(8)\% from \cite{hos10}.
The new evaluation of $P_{x}$ and $T_{1/2}$ for $\beta$-delayed
neutron emitters fostered by the IAEA~\cite{lia18} recommends a value
of $P_{1n} = 22.7(20)$\%, obtained by a weighted average of all measurements
except the first one.

$^{82}$Ga is the sole neutron emitter in its entire decay network.
The one-neutron emission window is $Q_{\beta 1n} = 5.290(3)$~MeV.
This value and the values of other decay energy windows in this paper are taken 
from the 2016 Atomic Mass Evaluation \cite{hua17}.
The neutron energy spectrum has not been measured for this decay
but it was for the lighter isotopes $^{79-81}$Ga \cite{bra89}. From the evolution
of the shape one can deduce that most of the neutron spectrum for $^{82}$Ga
should be contained within 1~MeV. This is confirmed by theoretical calculations
of the delayed neutron spectrum, which can be retrieved from the ENDF/B-VII.1 
data base \cite{endfb71}.
This spectrum is calculated from $\beta$-strength distributions obtained 
within the QRPA formalism \cite{moe03} and neutron emission rates obtained 
within the Hauser-Feshbach formalism \cite{kaw08}. 
Thus we conclude that the use of the nominal
neutron efficiency $\bar{\varepsilon}_{n} = 66.8 (20)\%$ 
(Section~\ref{sec:setup}) should be appropriate in this decay.

\begin{figure}[ht]
 \begin{center}
 \includegraphics[width=7.8cm]{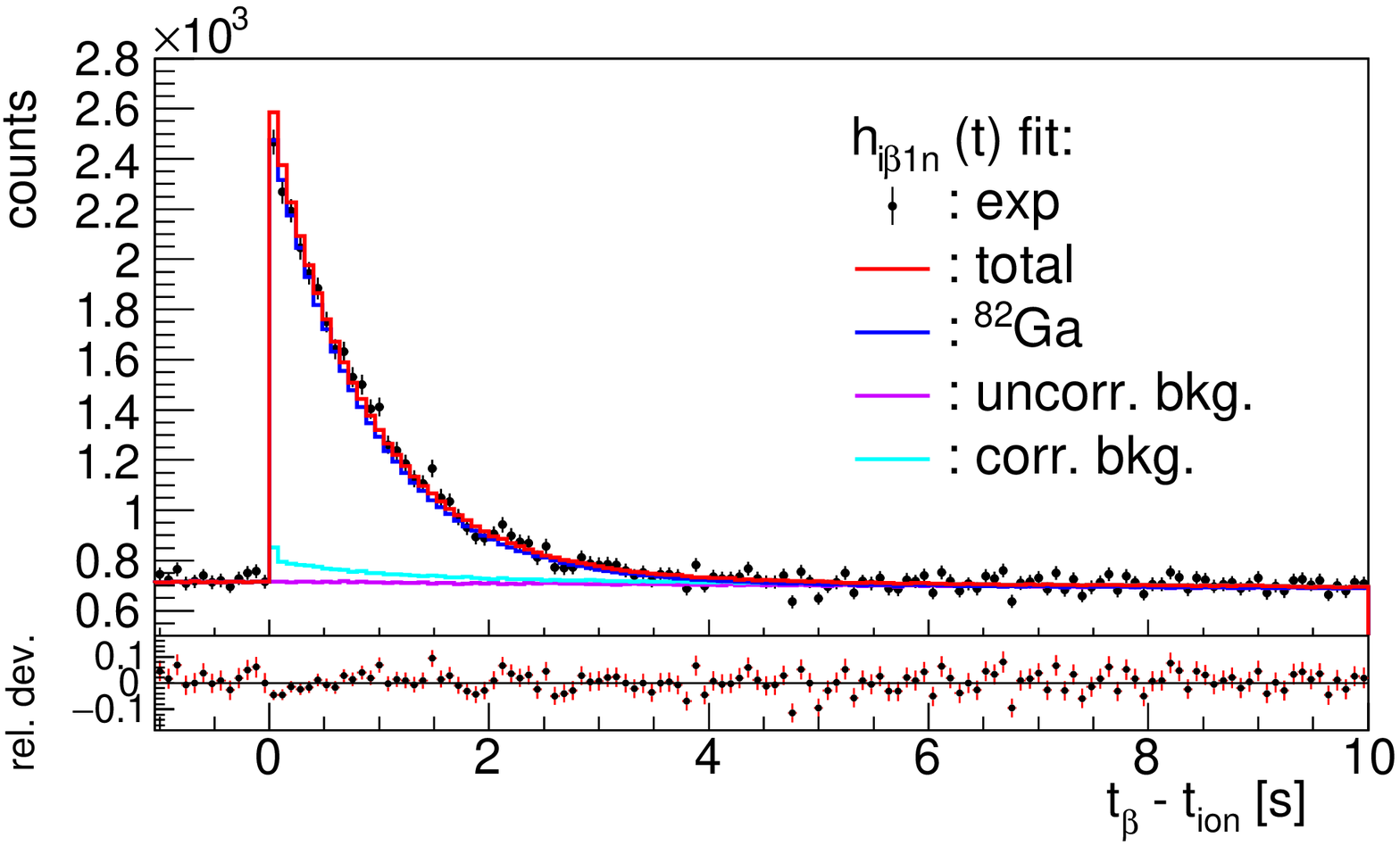}
 \includegraphics[width=7.8cm]{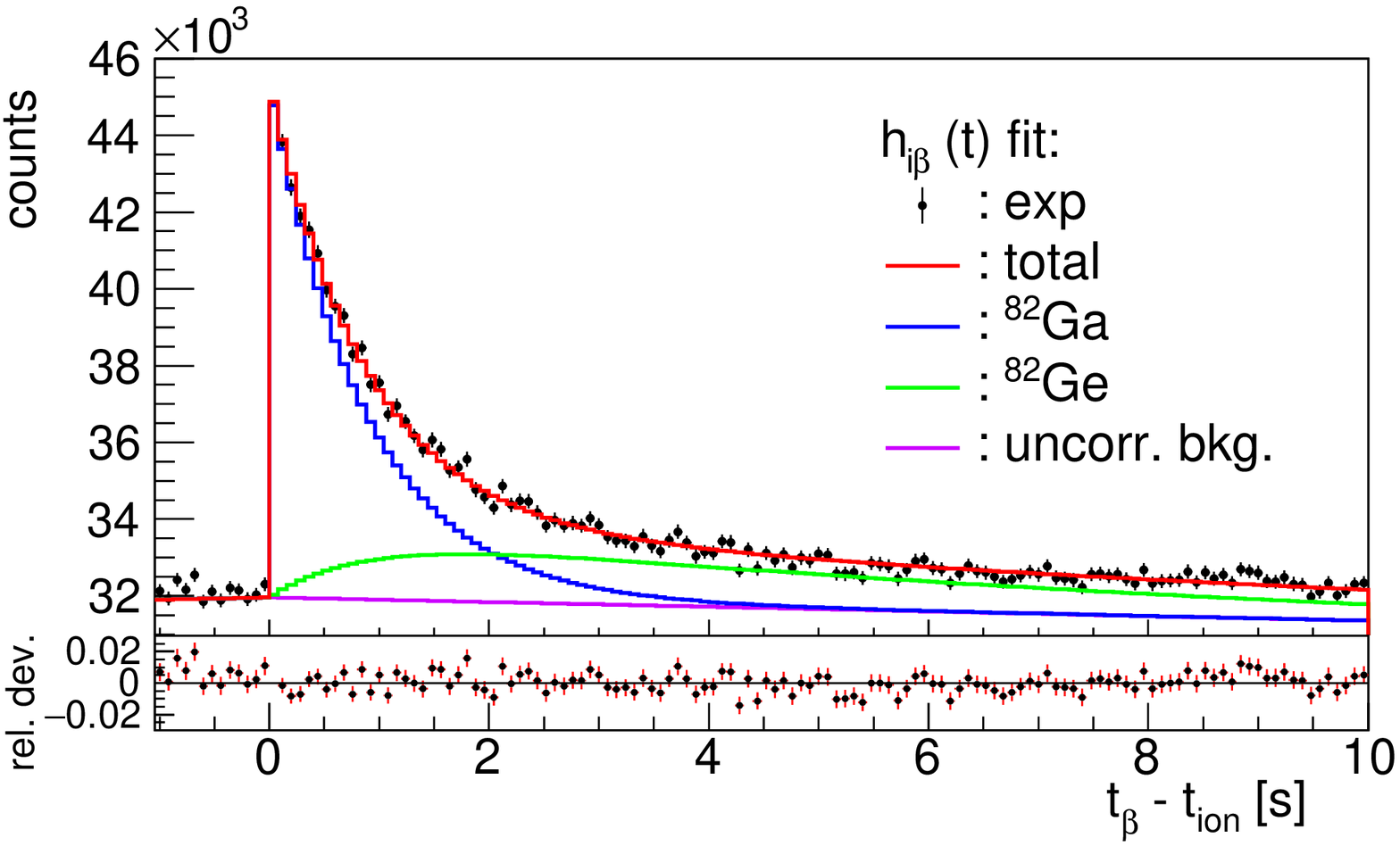}
 \end{center}
 \caption{Fit to implant-$\beta$ (bottom panel) and implant-$\beta$-1n (top panel) 
time correlation histograms for the decay of $^{82}$Ga. In both panels the red line
represents the total fit function, the violet line the uncorrelated background 
and the blue line the contribution of parent decay. In addition  the green line in the bottom
panel represents the daughter contribution, and the light blue line in the top panel the correlated
neutron background contribution. Additional smaller descendant contributions to the bottom panel
are not shown for clarity. The relative deviation of the data with respect to the fit is shown
in lower part of each panel.}
 \label{fig:fit82ga}
\end{figure}

Our data and the fits are shown in Fig.~\ref{fig:fit82ga}. 
There is a large difference between the fit and measurement 
at the first positive time bin in the $h_{i\beta} (t)$ histogram 
(with $6 \times 10^{4}$ counts it is outside the range shown).
This is caused
by the ion-induced $\beta$ background.
The corresponding histogram bin is not included in the fit region.
The number of accidental one-neutron counts per detected $\beta$ 
is $r_{1} = 0.006793(29)$, much smaller than the values obtained
in the May-June 2017 run (see Section \ref{sec:background}). 
This reflects the different background conditions
in the two experiments.
In the fit all the decay branches down to stable
nuclei are followed. The half-life of all descendants is relatively well known \cite{ensdf}.
The half-life of $^{82}$Ga is also well known, $T_{1/2}=601(2)$~ms \cite{lia18}, 
and was fixed in the fit.
Ambiguities appear in the case of $^{81}$Ge with two known $\beta$-decaying
isomers.  However, both  
have equal half-life within uncertainties according to \cite{hof81},
which minimizes the impact of the respective unknown population.
Also in the case of $^{82}$As two isomers are known
but the decay of the $0^{+}$ ground state of $^{82}$Ge
will populate weakly the $(5^{-})$ isomer and it was neglected.
The case of $^{81}$Se, again with two isomers, poses no problem since  
it contributes marginally to the decay activity.

The decay energy window for $^{82}$Ga is large, $Q_{\beta} = 12.484(3)$~MeV, 
much larger than the $Q_{\beta}$ for other contributing decays. In particular
it is nearly 8~MeV larger than the $Q_{\beta}$ of the daughter  $^{82}$Ge,
the second largest contributor in the decay chain.
Therefore the fit was performed including as a free parameter the $\beta$ efficiency 
for the parent decay, resulting in a value of $P_{1n} = 19.10(46)$\%. 
If we keep all $\beta$ efficiencies fixed the 
fit is poorer ($\chi^{2}/\nu= 1.3$ instead of 0.98, see also Fig.~\ref{fig:fitdev}) 
and the result becomes 13\% larger $P_{1n} = 21.60(30)$\%.
The fitted $\beta$ efficiency is $84.6(20)$\%, relative to the efficiency
for the remaining decay branches, that are kept fixed.
This can be interpreted in the light of the simulations presented
in Fig.~\ref{fig:beff} showing that if on average the decay
proceeds by large decay energies the efficiency can be lower
than if it decays with smaller $\beta$ energies.

The uncertainty on $P_{1n}$ values quoted above is obtained from the fit
and represents the statistical uncertainty. 
The systematic uncertainty
due to the uncertainties in the half-life of the parent and all descendants was evaluated as 0.16\%
using the parameter sampling procedure described before.
The uncertainty due to the assumed uncertainty in the neutron efficiency
amounts to 0.61\%. The uncertainty due to the background corrections (correlated
and uncorrelated) is evaluated as 0.18\%. The total systematic uncertainty is 0.65\%.
Combining quadratically the statistical and systematic uncertainties our result is 
$P_{1n} = 19.10(80)$\%. It agrees within uncertainties with the weighted average 
of previous measurements with the lower $P_{1n}$ values  \cite{lun80,war86,tes16} ,
$P_{1n} = 20.4(12)$\%. The result confirms also the value of the nominal
efficiency used.

\subsection{$^{80}$Zn: sensitivity limit to small $P_{1n}$}
\label{sec:80zn}

The importance of a proper correction of accidentally correlated neutron background 
in the case of  
weak two-neutron emitters was demonstrated in Section \ref{sec:background}.
$^{80}$Zn with $Q_{\beta 1n} = 2.828(3)$~MeV and a small $P_{1n}$ value 
is a good example of the importance of background correction for weak one-neutron
emitters. It serves also as a test case to study 
the sensitivity limit of our experiment for $P_{1n}$ determination. 
There are two previous measurements of the delayed neutron emission probability
in $^{80}$Zn.
A rather uncertain value of 1.0(5)\% is reported in \cite{kra91} and an upper limit, 
$P_{1n} < 1.8$\%, is reported in \cite{hos10}. The new evaluation \cite{lia18}
adopts the value of \cite{kra91}.

\begin{figure}[ht]
 \begin{center}
 \includegraphics[width=7.8cm]{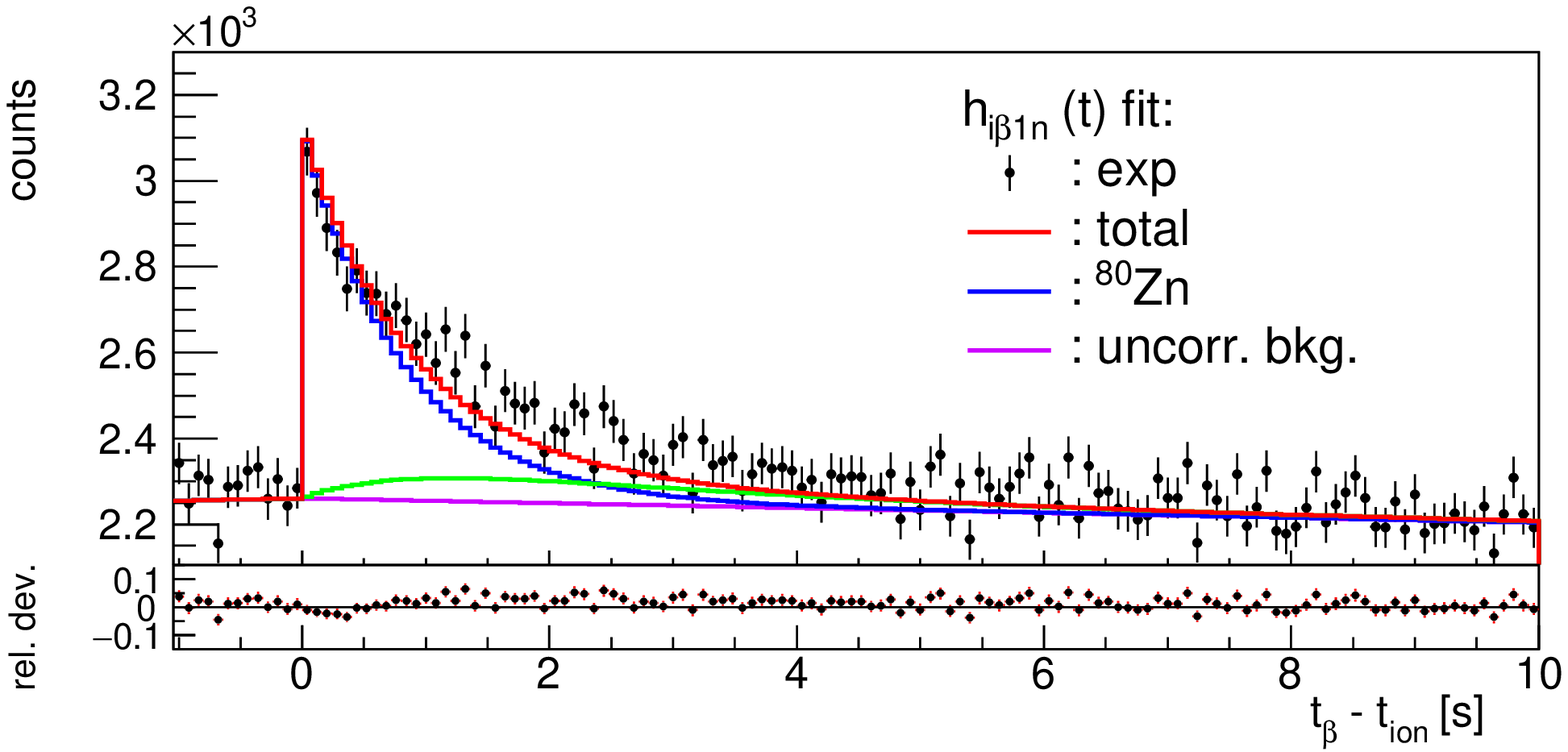}
 \includegraphics[width=7.8cm]{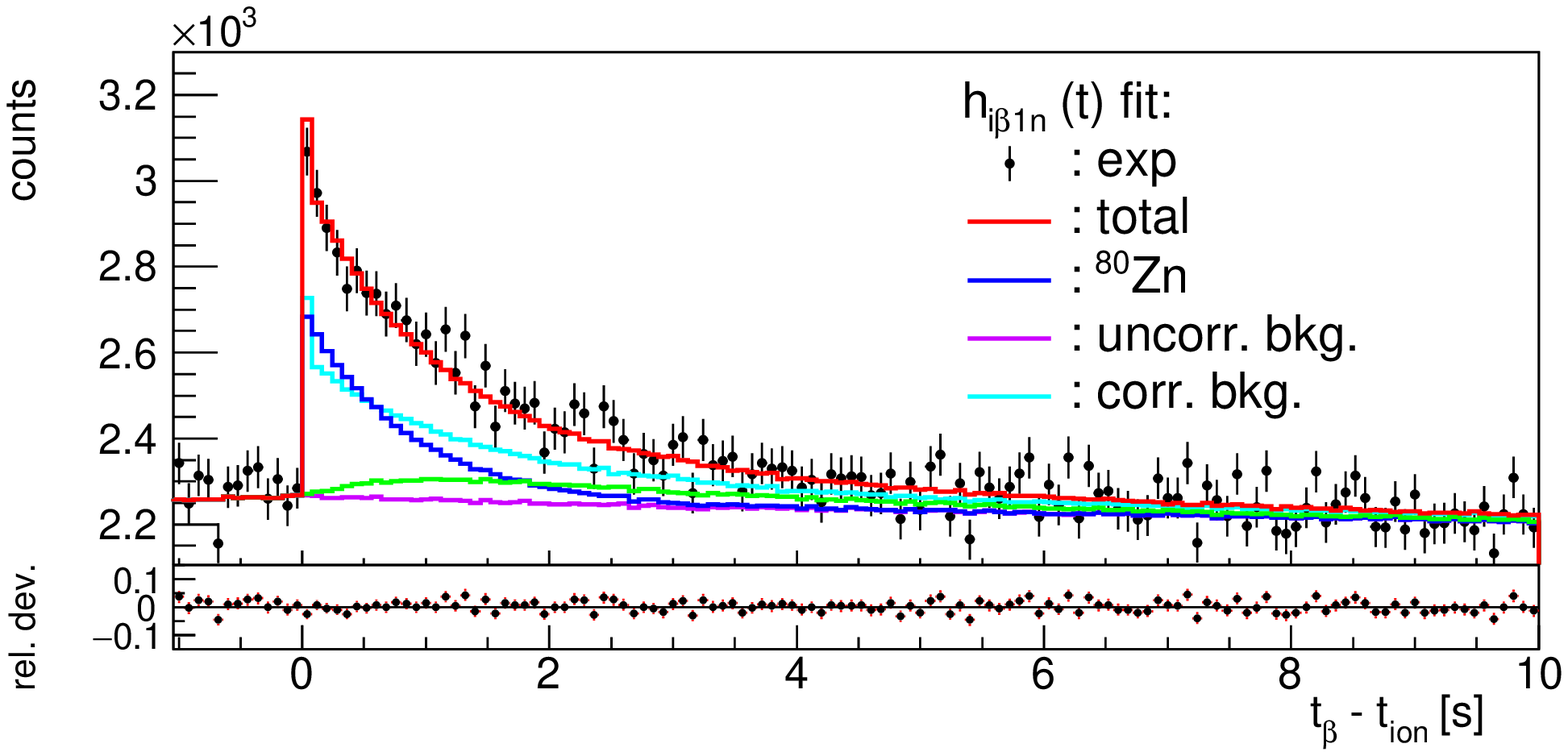}
 \includegraphics[width=7.8cm]{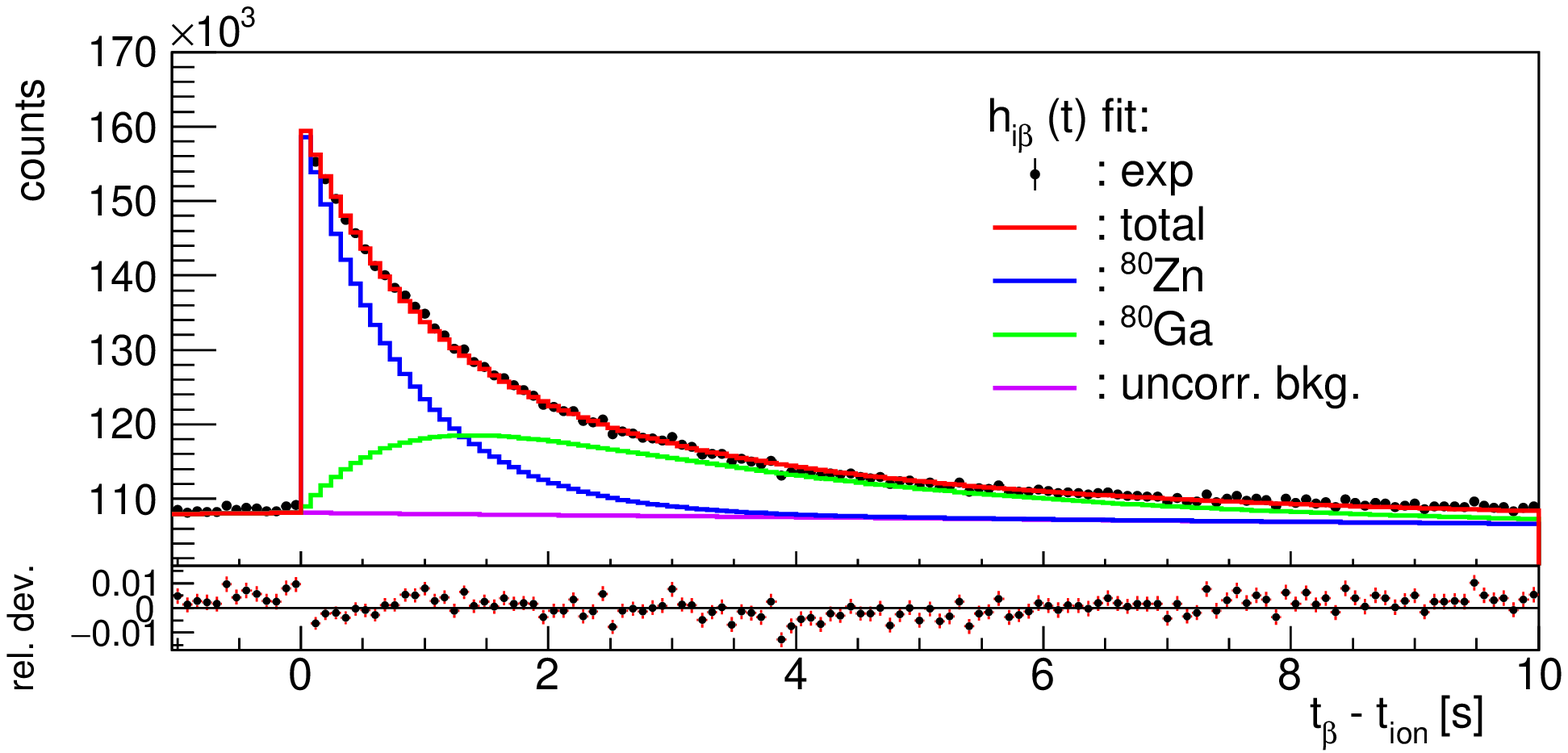}
 \end{center}			
 \caption{Fit to implant-$\beta$ (bottom panel) and implant-$\beta$-1n (top an central panels)
time correlation histograms for the decay of $^{80}$Zn. Top panel: fit  
without accidental one-neutron background correction.
Central panel: fit with accidental one-neutron background correction.
The same color code as in Fig.~\ref{fig:fit82ga} is used 
for the different contributions to the fit function.}
 \label{fig:fit80zn}
\end{figure}

One million $^{80}$Zn ions were implanted during the commissioning run.
In the decay chain \cite{lia18}, $^{80}$Ga is a known $\beta$-delayed neutron
emitter with a weak neutron branching $P_{1n} = 0.846(73)$\%
and $^{79}$Ga is an even weaker emitter with $P_{1n} = 0.084(29)$\%.
Two isomers are known \cite{ver13} in $^{80}$Ga 
with $T_{1/2}=1.3(2)$~s $(J_{\pi} = 3^{(-)})$ and $T_{1/2}=1.9(1)$~s $(J_{\pi} = 6^{(-)})$,
but the high-spin isomer is only weakly populated in the the decay of the $^{80}$Zn
$0^{+}$ ground state and it was ignored. 
The $Q_{\beta}$ of $^{80}$Zn, $7.575(4)$~MeV, is
actually smaller than that of the daughter $^{80}$Ga, $10.312(4)$~MeV,
at difference with the remaining cases shown in Fig.~\ref{fig:fitdev}.
Another characteristic of the decay of $^{80}$Ga is the sizable population of 
a $2^{+}$ state at $E_{x}= 659$~keV that emits conversion electrons and of a
$0^{+}$ state at $E_{x}= 639$~keV that can only decay
by electron conversion to the $0^{+}$ g.s. (E0 transition) \cite{got16}.
These low-energy conversion electrons are easily detected in AIDA
increasing the apparent $\beta$ efficiency for $^{80}$Ga decay.
Therefore we leave this efficiency as a free parameter of the fit.
The lower panel of Fig.~\ref{fig:fit80zn} shows the good quality
obtained in the fit to the $h_{i \beta} (t)$
histogram. The adjusted $\beta$ efficiency is 15\% larger
than the efficiency for other decay branches that are kept fixed.
This result can be understood as the effect of conversion electrons.

The upper panel of Fig.~\ref{fig:fit80zn} shows the fit to the implant-$\beta$-1n 
histogram without correction for the accidental correlated one-neutron 
background and the central panel the fit including the correction.
As can be seen this correction represents the largest contribution
to the measured histogram.
Without correction the fit to
$h_{i\beta 1n} (t)$ is poor because the correlated background has the shape of 
$h_{i\beta} (t)$ (see Section \ref{sec:background}) and the resulting $P_{1n} = 2.79(11)$\%
is too large. After correction we obtain $1.36(11)$\% in agreement with previous
results. A fit where all $\beta$ efficiencies are kept fixed results in a value of 1.28(11)\%. 
The systematic uncertainty due to background correction is 0.06\%.
The one due to the fixed parameters in the fit
(all $T_{1/2}$ and $P_{1/2}$ of descendants) is 0.02\% and that of the neutron efficiency
is 0.04\%. Combining all uncertainties our final result is $P_{1n} = 1.36(12)$\%.

This case shows also that rather accurate $P_{1n}$ values on the order of one percent
can be extracted from our data. This statement of course depends on the
implantation statistics. We have tested that analyzing one tenth of the present statistics
($10^{5}$ ions) one can still obtain a reasonable result of $P_{1n} = 1.62(47)$\%.

\subsection{$^{81}$Ga: sensitivity limit for small implant statistics}
\label{sec:81ga}

The decay window for neutron emission in $^{81}$Ga 
is $Q_{\beta 1n} = 3.836(3)$~MeV.
It is the only neutron emitter in the decay chain network.
There are three previous $P_{1n}$  measurements that agree relatively well: 
12.0(9)\% \cite{lun80}, 10.6(8)\% \cite{war86}, and 12.9(8)\% \cite{rud93}.
The new evaluation \cite{lia18} recommends the value of $P_{1n} = 12.5(8)$\%.
The half-life is also well known $T_{1/2} = 1.217(4)$~s.
The number of implanted ions during the commissioning run was 4400.
Thus this case serves to test the sensitivity limit of our setup with low statistics.

\begin{figure}[ht]
 \begin{center}
 \includegraphics[width=7.8cm]{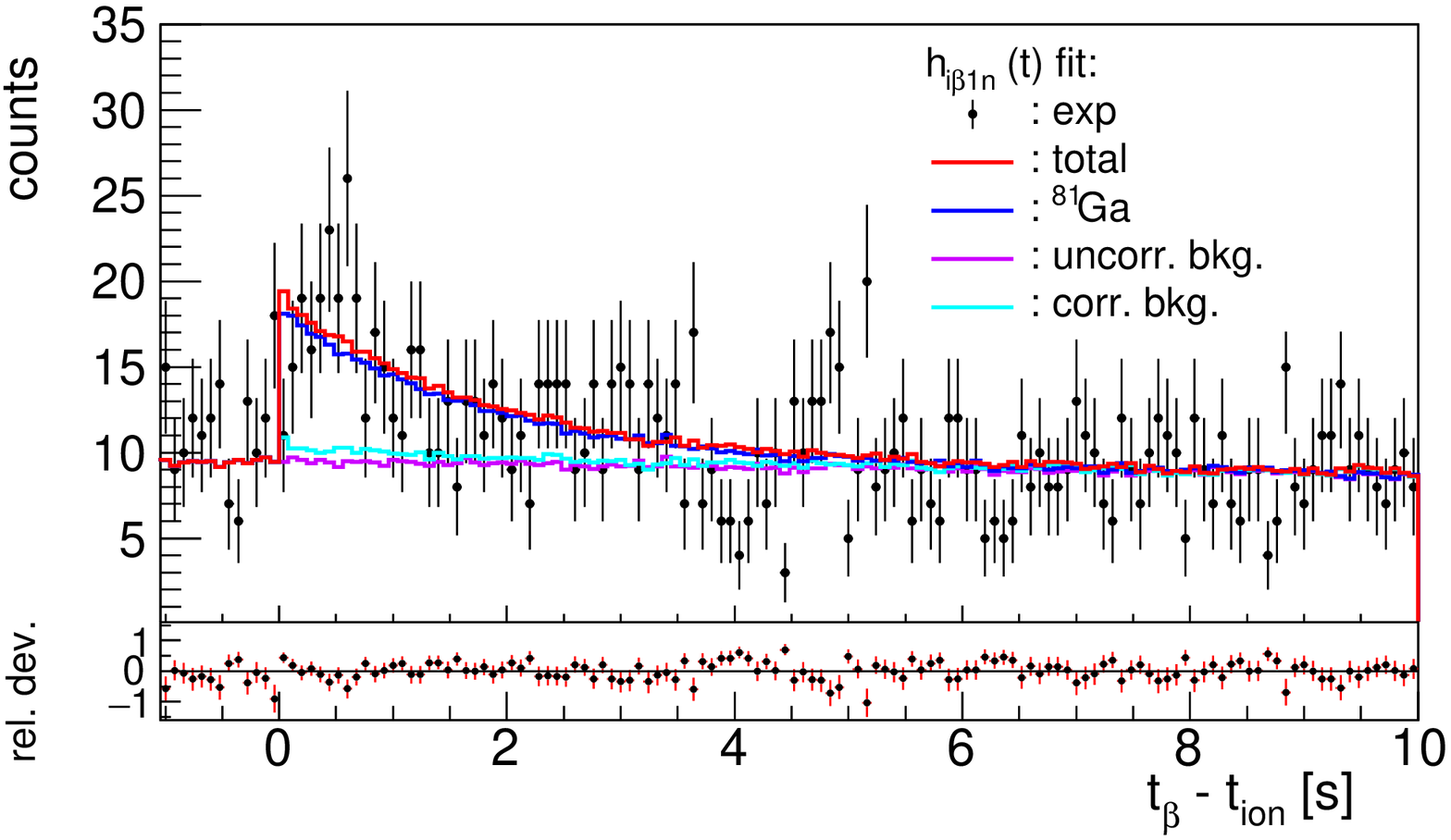}
 \includegraphics[width=7.8cm]{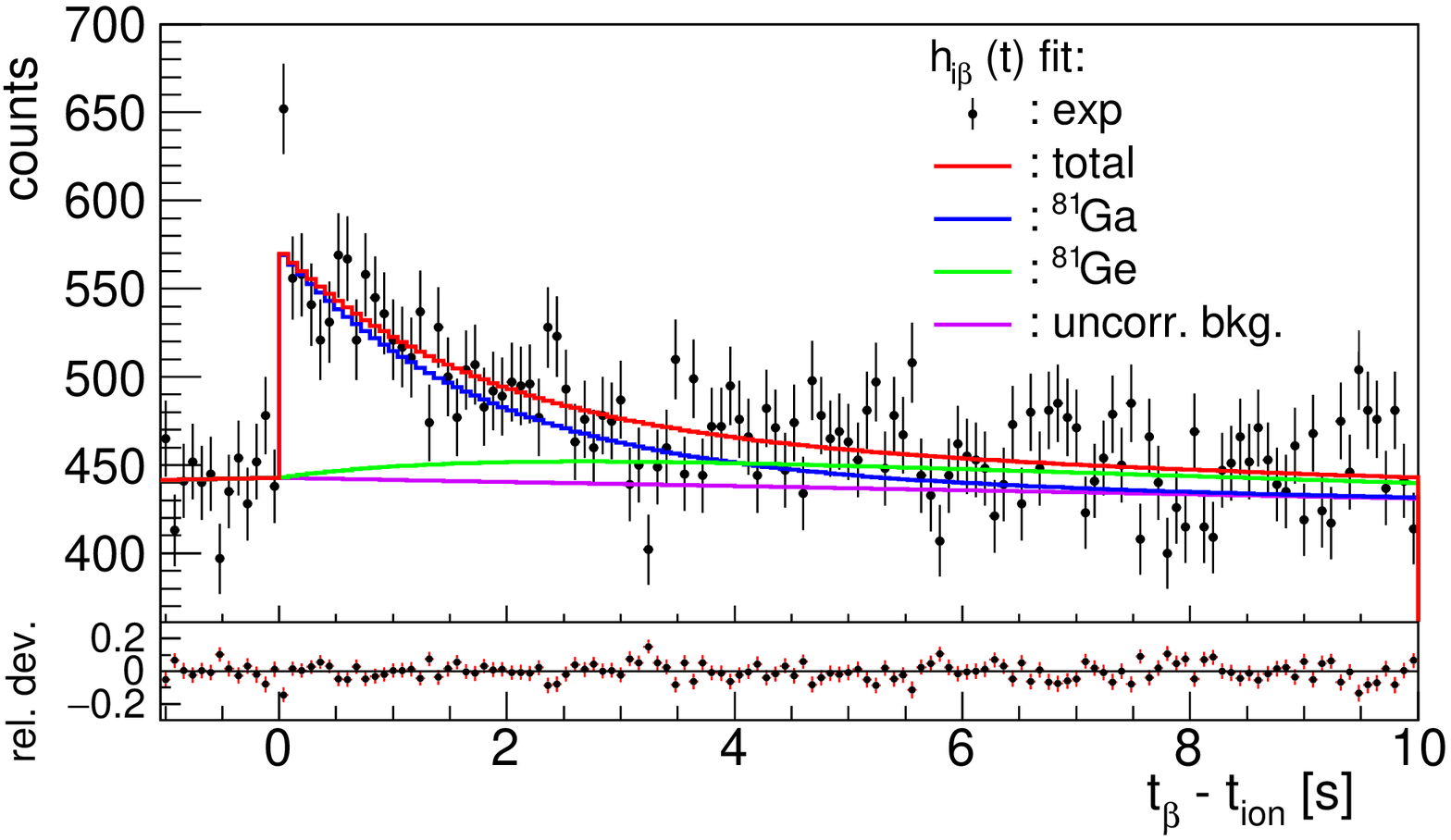}
 \end{center}
 \caption{Fit to implant-$\beta$ (bottom) and implant-$\beta$-1n (top) 
time correlation histograms for the decay of $^{81}$Ga. The same color code
as in Fig.~\ref{fig:fit82ga} is used for the different contributions to the fit function.}
 \label{fig:fit81ga}
\end{figure}

Figure \ref{fig:fit81ga} shows the result of the fit using the nominal neutron efficiency
and fixing the half-life of descendants to the values in the ENSDF database \cite{ensdf}.
The $\beta$ efficiencies were kept fixed during the fit.
A total of 190 neutrons stand out from a background of 820 neutrons. 
The $P_{1n}$ from the fit is 11.3(23)\%. 
The systematic uncertainty due to parameters that are kept fixed in the fit is 1.2\%.
Our result is then $P_{1n}=11.2(26)$ in agreement with previous results.

This demonstrates that with a few thousand ions we are able to measure
$P_{1n}$ values of the order of 5-10\% with accuracies in the order of 25\%.

\section{Conclusion}
\label{sec:conclusion}

We have carried out the commissioning of a new setup for the measurement of decay properties
of $\beta$-delayed neutron emitters using radioactive beams at RIKEN.
This allowed us to verify the performance of the BRIKEN neutron counter under
experimental conditions. We found that the beam induced neutron background 
in the detector is about 2-3 orders of magnitude larger than the natural neutron background.
The background rate is quite sensitive to the spectrometer setting. Minimizing the material
in the beam path close to the detector helps to reduce the background. We found that
another effective way of reducing the background is to veto neutron signals coming
shortly after any beam particle enters the experimental area. 
This reduces the one-neutron background rate by a factor 2-3 and the two-neutron 
background rate by a factor $\sim$30. The large background imposes a limit on the
minimum measurable $P_{1n}$ that otherwise depends on the statistics (number of
implanted ions) and the value of $P_{1n}$ itself. We demonstrated that we are able
to determine $P_{1n}$ values of the order of 1\% with $10^{5}$ ions. 
We could determine also $P_{1n}$ values of the order of 10\% with
few thousand  implanted ions.
For $P_{2n}$ values the situation is more favorable because of the
much higher background reduction.

Because of the large size of the $\beta$-neutron coincidence window (200~$\mu$s)
the number of accidental $\beta$-neutron correlations is large.
This introduces a distortion of the $\beta$-implant-neutron time correlation spectra
that severely affects the determination of small $P_{1n}$ and $P_{2n}$ values.
We have introduced a novel method, based on time-reversed correlations,
to determine this distortion accurately and correct for it.

Systematic errors due to the unknown dependence of $\beta$ and neutron efficiency
on nucleus and decay branch have been discussed. Although the design of the BRIKEN
neutron counter minimizes the neutron energy dependence of the efficiency,
reductions of up to about 10\% with respect to the nominal neutron efficiency can be expected
for decays with large $Q_{\beta 1n}$ windows. The correct efficiency can be calculated
from the neutron spectrum and simulated efficiencies. In cases where the spectrum
is unknown one can use theoretical estimates to compute the correction factor.
Another possibility, that we are currently investigating, 
is to use the number of counts per
detector ring, which is sensitive to the neutron energy distribution, 
to determine directly the effective average efficiency.

The evaluation of systematic errors due to differences in $\beta$ efficiencies is more challenging. 
As this effect is related to the threshold in the $\beta$ detector, minimization 
of the threshold for $\beta$ events in the sorting of AIDA data is a requisite for 
accurate $P_{xn}$ determinations.
This is a demanding task given the complexity of this type of detector
and the varying conditions in different experiments.
Currently we are actively working to improve the $\beta$ event reconstruction in AIDA.
For the commissioning run we selected a sort that is a compromise between
threshold reduction and signal-to-noise ratio. 
We found evidence  for a residual $\beta$  efficiency effect in the
fits to these data. These are in general cases where the 
parent decay $Q_{\beta}$ is quite large, much larger than
the decay energy window for other contributing decay branches.
Our approach to solve this issue is to include the parent decay
$\beta$ efficiency as an adjustable parameter in the fit.
We also studied a case where the $\beta$ efficiency of the daughter
decay is increased due to the emission of conversion electrons.

We presented the result of the analysis for few selected cases 
measured in the commissioning run. 
The results obtained for other cases will be presented in a forthcoming publication.
They confirm the value of the neutron efficiency for the current setup. 
They show also the importance of an accurate correction
of the correlated neutron background. 
In general they confirm the good performance of the detector
setup and the expected quality of the results 
from the experiments that have been already performed or are planned .

\section*{Acknowledgment}
This work has been supported by the Spanish Ministerio de Econom\'{\i}a
y Competitividad under grants FPA2011-24553, FPA2011-28770-C03-03,
FPA2014-52823-C2-1/2, IJCI-2014-19172 and SEV-2014-0398, 
and  by FP7/EURATOM Contract No. 605203.
This work has been supported by the Office of Nuclear Physics, U. S.
Department of Energy under the contract DE-AC05-00OR22725 (ORNL).
This research was sponsored in part by the Office of Nuclear Physics,
U.S. Department of Energy under Award No. DE-FG02-96ER40983 and by the National
Nuclear Security Administration under the Stewardship Science Academic Alliances program
through DOE Award No. DE-NA0002132.
This work was supported by JSPS KAKENHI
(Grants Numbers 14F04808, 17H06090, 25247045, 19340074).
Supported by the UK Science and Technology Facilities Council.
This work has been supported by the Natural Sciences and Engineering
Research Council of Canada (NSERC) via the Discovery Grants SAPIN-2014-00028 and
RGPAS 462257-2014.
This work has been supported by the National Science Foundation grant
PHY 1714153.
Supported by the National Science Foundation under Grant Number PHY-
1430152 (JINA Center for the Evolution of the Elements), Grant Number PHY-1565546 (NSCL).
Supported by the UK Science and Technology Facilities Council grant
No. ST/N00244X/1.
This work was supported by NKFIH (K120666).
Supported by South Korea NRF grants 2016K1A3A7A09005575 and 2015H1A2A1030275.
Work partially done within IAEA-CRP for Beta Delayed Neutron Data.
G. G. Kiss acknowledges support form the Janos Bolyai research fellowship.
This experiment was performed at RI Beam Factory operated by RIKEN
Nishina Center and CNS, University of Tokyo.



\appendix
\section{}
\label{sec:appendix}

We describe in this Appendix how to  obtain background
correction formulas for the analysis of BRIKEN data from $\beta$2n emitters.
We calculate the effect of accidental
coincidences with background neutrons on the measured histograms 
$h_{i\beta 1n} (t)$ and $h_{i\beta 2n} (t)$ that, together with $h_{i\beta}$, 
are needed to obtain $P_{1n}$, $P_{2n}$ and $T_{1/2}$. 
At the end of the Appendix we give also,
without deduction, the corresponding formulas for the case of  $\beta$3n emitters
which can be obtained following a similar line of reasoning.

Let us consider first  the  $h_{i\beta 2n} (t)$ histogram.
The corresponding unperturbed time distribution is represented 
by the function $f_{i\beta 2n} (t)$ defined in Eq.~\ref{eq:fib2n}.
As mentioned in Section \ref{sec:background}, 
one of the effects of accidental coincidences with background neutrons
is the loss of counts, resulting in a scaling of this distribution
of the form $(1-r) f_{i\beta 2n} (t)$, where $r$ is the total probability of 
accidental neutron coincidences per $\beta$.
In addition to this effect accidental coincidences produce spurious counts 
in the histogram that come from three different sources. 

The first contribution comes from $\beta$ particles
that do not see correlations with decay neutrons
but accidentally correlate with two background neutrons coming within
the $\Delta t_{\beta n}$ coincidence window. The probability of accidental
correlation with two bacground neutrons is given by
$r_{2}$, see Eq.~\ref{eq:r2}.
The time distribution of $\beta$ events that do not see
correlations with decay neutrons can be obtained by subtraction from the 
distribution of all $\beta$ events, $f_{i\beta} (t)$ (Eq.~\ref{eq:fib}),
those events where decay neutrons are detected. 
For the $\beta$1n decay channel this is represented
by  $f_{i\beta 1n} (t)$ (Eq.~\ref{eq:fib1n}). For the 
$\beta$2n decay channel, two terms appear.
The first term corresponds to events where the two neutrons
are detected and is represented by $f_{i\beta 2n} (t)$. The second
term corresponds to events where only one of the two neutrons is detected.
The probability of detecting only one of the two neutrons emitted is given 
by  $2 (1-\bar{\varepsilon}_{2n}) \bar{\varepsilon}_{2n}$,
assuming that the neutron detection efficiency for both neutrons
in the $\beta$2n channel (Eq.~\ref{eq:eff2n}) is equal. 
Taking into account the dependence of $f_{i\beta 2n} (t)$ (Eq.~\ref{eq:fib2n})
with $\bar{\varepsilon}_{2n}$ then  $2 r_{e} f_{i\beta2n} (t))$,
with $r_{e} = (1-\bar{\varepsilon}_{2n}) / \bar{\varepsilon}_{2n}$,
represents the distribution of $\beta2n$ events where only one neutron is detected.
Taking both terms into consideration, this contribution takes the form
$ r_{2} (f_{i\beta} (t) - f_{i\beta 1n} (t) - f_{i\beta 2n} (t) - 2 r_{e} f_{i\beta 2n} (t))$.

The second contribution comes from events 
belonging to the $\beta1n$ decay channel
when in addition to the detection of a decay neutron, 
with time distribution given by $f_{i\beta 1n} (t)$,
a background neutron accidentally arrives within $\Delta t_{\beta n}$
with probability $r_{1}$ (Eq.~\ref{eq:r1}). 
This results  in a term of the form $r_{1}f_{i\beta 1n} (t)$.

The third contribution, analogous to the second one, 
comes from the $\beta2n$ channel itself when one
of the two neutrons emitted escapes detection (see above) 
but a single background neutron
arrives accidentally within $\Delta t_{\beta n}$ with probability $r_{1}$. 
This gives a contribution of the form $2 r_{e} r_{1}f_{i\beta 1n} (t)$

The measured histogram is the sum off all these contributions
plus the uncorrelated background $h_{i u \beta 2n} (t)$ (see Section~\ref{sec:background}):

\begin{equation}
\begin{split}
h_{i\beta 2n} (t) = &  (1-r) f_{i\beta 2n} (t) \\
 & + r_{2} (f_{i\beta} (t) - f_{i \beta 1n} (t) - f_{i\beta 2n} (t) - 2 r_{e} f_{i\beta 2n} (t))\\
 & + r_{1} (f_{i \beta 1n} (t) +2 r_{e} f_{i\beta 2n} (t)) \\
 & + h_{i u \beta 2n} (t)
\label{eq:a1}
\end{split}
\end{equation}

Let us consider now the $h_{i\beta 1n} (t)$ histogram.
In the case of $\beta2n$ emitters one needs to modify
Eq.~\ref{eq:c1b1n} describing
the relation between the measured histogram 
and the unperturbed time distributions. 
The term representing the loss of events by accidental coincidences
with background neutrons remains the same, $(1-r) f_{i\beta 1n} (t)$.
The  term representing spurious counts coming from accidental correlations
with single background neutrons, with probability $r_{1}$,
needs to be modified. As explained above the time distribution
of events that do not see correlations with decay neutrons
must take into account the contributions of the $\beta$2n channel.
The term takes the form 
$ r_{1} (f_{i\beta} (t) - f_{i\beta 1n} (t) - f_{i\beta 2n} (t) - 2 r_{e} f_{i\beta 2n} (t))$.
In addition, one needs to consider a new term contributing
to the spurious counts coming from the
the $\beta$2n channel when only one
of the two neutrons is detected, represented by the distribution 
$2 r_{e} f_{i\beta 2n} (t)$,
and there is no accidental coincidence with background neutrons,
with a probability $1-r$. 
Thus this contribution takes the form $2 r_{e} (1-r) f_{i\beta2n} (t)$.
With these modifications and including the uncorrelated background
contribution $h_{i u \beta 1n} (t)$, the measured $h_{i\beta1n} (t)$
histogram can be evaluated as

\begin{equation}
\begin{split}
h_{i\beta 1n} (t) = & (1-r) f_{i\beta 1n} (t) \\
& + r_{1} (f_{i\beta} (t) - f_{i\beta 1n} (t) - f_{i\beta 2n} (t) - 2 r_{e} f_{i\beta 2n} (t))\\
& + 2 r_{e} (1-r) f_{i \beta 2n} (t) \\
& + h_{i u \beta 1n} (t)
\label{eq:a2}
\end{split}
\end{equation}

Rearranging terms in Eq. \ref{eq:a1} and  Eq. \ref{eq:a2} 
the relation between the measured histograms and the 
unperturbed distributions can be written down in a compact form:

\begin{equation}
\begin{split}
h_{i\beta} (t) - h_{iu\beta} (t)& =  f_{i \beta} (t) \\ 
h_{i\beta 1n} (t) - h_{iu\beta 1n} (t) & =  a_{0} f_{i \beta} (t) + a_{1} f_{i \beta 1n} (t) + a_{2} f_{i \beta 2n} (t) \\ 
h_{i\beta 2n} (t) - h_{iu\beta 2n} (t)& =  b_{0} f_{i \beta} (t) + b_{1} f_{i \beta 1n} (t) + b_{2} f_{i \beta 2n} (t) 
\label{eq:a3}
\end{split}
\end{equation}

The coefficients appearing in this formula are given by

\begin{equation}
\begin{split}
a_{0} & = r_{1} \\
a_{1} & = 1 -r -r_{1} \\
a_{2} & = 2 r_{e} (1-r - r_{1}) -r_{1} \\
b_{0} & = r_{2} \\
b_{1} & =  r_{1}-r_{2} \\
b_{2} & =  1 -r -r_{2} + 2 r_{e} (r_{1}-r_{2})
\label{eq:a4}
\end{split}
\end{equation}

If we denote with $h'_{i\beta} (t)$, $h'_{i\beta 1n} (t)$ 
and $h'_{i\beta 2n} (t)$, the histograms
corrected for the uncorrelated background 
($h'_{i\beta} (t)= h_{i\beta} (t) - h_{iu\beta} (t), ...)$, 
and substitute $f_{i \beta} (t)$ for $h'_{i\beta} (t)$ 
in the two lower rows of equation \ref{eq:a3}, one can solve 
this system of equations for $f_{i \beta 1n} (t)$ and $f_{i \beta 2n} (t)$:

\begin{equation}
\begin{split}
f_{i \beta 1n} (t) & = d_{0} h'_{i\beta} (t) + d_{1} h'_{i\beta 1n} (t) + d_{2} h'_{i\beta 2n} (t)\\
f_{i \beta 2n} (t) & = e_{0} h'_{i\beta} (t) + e_{1} h'_{i\beta 1n} (t) + e_{2} h'_{i\beta 2n} (t)
\label{eq:a5}
\end{split}
\end{equation}

with the coefficients  given by

\begin{equation}
\begin{split}
d_{0} & =  -\frac{a_{0} b_{2}-b_{0} a_{2}}{a_{1} b_{2} - b_{1} a_{2}} \\
d_{1} & =  \frac{b_{2}}{a_{1} b_{2} - b_{1} a_{2}} \\
d_{2} & =  -\frac{a_{2}}{a_{1} b_{2} - b_{1} a_{2}} \\
e_{0} & =  -\frac{b_{0}a_{1}-a_{0} b_{1}}{a_{1} b_{2} - b_{1} a_{2}} \\
e_{1} & =  -\frac{b_{1}}{a_{1} b_{2} - b_{1} a_{2}} \\
e_{2} & =  \frac{a_{1}}{a_{1} b_{2} - b_{1} a_{2}}
\label{eq:a6}
\end{split}
\end{equation}

One can interpret Eq.~\ref{eq:a5} as representing the corrected
histograms $h^{corr}_{i\beta 1n}$ and $h^{corr}_{i\beta 2n}$
respectively. Alternatively they can be used to obtain the form of 
the fit functions  for the measured histograms
including all background components:

\begin{equation}
\begin{split}
h_{i \beta 1n} (t) & = -\frac{d_{0}}{d_{1}} h_{i\beta} (t) + 
\frac{1}{d_{1}} f_{i\beta 1n} (t) - \frac{d_{2}}{d_{1}} h_{i\beta 2n} (t) + \tilde{h}_{iu \beta 1n} (t)\\
h_{i \beta 2n} (t) & = -\frac{e_{0}}{e_{2}} h_{i\beta} (t) - 
\frac{e_{1}}{e_{2}} h_{i\beta 1n} (t) + \frac{1}{e_{2}} f_{i\beta 2n} (t) + \tilde{h}_{iu \beta 2n} (t)
\label{eq:a7}
\end{split}
\end{equation}

Here $\tilde{h}_{iu \beta 1n} (t)$ and $\tilde{h}_{iu \beta 2n} (t)$
represent the remaining uncorrelated background after subtraction
of the scaled correction histograms.

For the case of a three-neutron emitter 
a similar line of reasoning gives
the relation between the measured histograms
$h_{i\beta} (t)$ and $h_{i\beta xn} (t)$,
corrected for the uncorrelated background contribution,
and the unperturbed distributions $f_{i\beta} (t)$ and  $f_{i\beta xn} (t)$. 
In particular it should take into account the contribution of the $\beta$3n decay
channel to the one-neutron and two-neutron time correlation histograms.
For convenience we give here, without deduction, the result: 

\begin{equation}
\begin{split}
h'_{i\beta} (t) & =  f_{i \beta} (t) \\ 
h'_{i\beta 1n} (t) & =  a_{0} f_{i \beta} (t) + a_{1} f_{i \beta 1n} (t) 
+ a_{2} f_{i \beta 2n} (t) + a_{3} f_{i \beta 3n} (t) \\ 
h'_{i\beta 2n} (t) & =  b_{0} f_{i \beta} (t) + b_{1} f_{i \beta 1n} (t) + b_{2} f_{i \beta 2n} (t) + b_{3} f_{i \beta 3n} (t) \\
h'_{i\beta 3n} (t) & =  c_{0} f_{i \beta} (t) + c_{1} f_{i \beta 1n} (t) + c_{2} f_{i \beta 2n} (t) + c_{3} f_{i \beta 3n} (t) 
\label{eq:a8}
\end{split}
\end{equation}

The coefficients $a_{0}$, $a_{1}$, $a_{2}$, $b_{0}$, $b_{1}$, and $b_{2}$, 
are identical to those given in Eq.~\ref{eq:a4}, 
and the new coefficients that appear are given by

\begin{equation}
\begin{split}
a_{3} & = 3 r^{2}_{e} (1-r - r_{1}) -(1 + 3 r_{e}) r_{1} \\
b_{3} & = 3 r_{e} (1 -r  - r_{2}) + 3 r_{e}^{2} (r_{1} - r_{2}) - r_{2}\\
c_{0} & = r_{3} \\
c_{1} & =  r_{2}-r_{3} \\
c_{2} & =  r_{1} - r_{3} + 2 r_{e} (r_{2} - r_{3}) \\
c_{3} & =  1 -r -r_{3} + 3 r_{e} (r_{1} - r_{3}) + 3 r_{e}^{2} (r_{2} - r_{3})
\label{eq:a9}
\end{split}
\end{equation}




\section*{References}

\begin{thebibliography}{00}


\bibitem{kod73} 
T. Kodama and K. Takahashi, Phys. Lett. B 43 (1973) 167.

\bibitem{mum16}
M. R. Mumpower et al., Prog. Part. Nucl. Phys. 86 (2016) 86.

\bibitem{tai18} 
J. L. Tain et al., Acta Phys. Pol. B 49 (2018) 417.

\bibitem{crpdn}
IAEA CRP on a reference data base for beta-delayed neutron emission,
https://www-nds.iaea.org/beta-delayed-neutron/

\bibitem{tar17}
A. Tarife\~{n}o-Saldivia et al., J. Instrum. 12 (2017) P04006.

\bibitem{gri17}
C. J. Griffin et al., Jap. Phys. Soc. Conf. Proc. 14 (2017) 020622.


\bibitem{oku12}
H. Okuno et al., Prog. Theor. Exp. Phys. 2012, 03C002.

\bibitem{kub12}
T. Kubo et al., Prog. Theor. Exp. Phys. 2012, 03C002.

\bibitem{aida}
https://www2.ph.ed.ac.uk/~td/AIDA/

\bibitem{nis13}
S. Nishimura et al., RIKEN Accelerator Progress Report 46 (2013) 182.

\bibitem{grz18} R. Grzywacz et al., accepted to RIKEN Accel. Prog. Rep. 51 (2018).

\bibitem{als15} 
M. Alshudifat et al., Physics Procedia 66 ( 2015) 445

\bibitem{gom11}
M. B. Gomez-Hornillos et al., J. Phys. Conf. Ser. 312 (2012) 052008.

\bibitem{grz14}
R. Grzywacz et al., Acta Phys. Pol. B 45 (2014) 217.

\bibitem{GE}
https://www.gepower.com/

\bibitem{LND}
http://www.lndinc.com/

\bibitem{geant4}
S. Agostinelli et al., Nucl. Instrum. Meth. Section A 506 (2003) 250.

\bibitem{bra89}
M. C. Brady, Evaluation and application of delayed neutron precursor data, 
PhD Thesis, Texas A\&M Univesity, LANL Thesis Report LA-11534-T (1989).

\bibitem{endfb71}
ENDF/B-VII.1 Evaluated Nuclear Data Library, http://www.nndc.bnl.gov/endf/b7.1/

\bibitem{man87}
W. Manhart, Cf-252 Neutron Spectrum, Report IAEA-NDS-98, 1987. 

\bibitem{tar18}
A. Tarife\~{n}o-Saldivia et al., to be published.

\bibitem{mpr16}
https://www.mesytec.com/

\bibitem{mpod}
http://www.wiener-d.com/

\bibitem{gro00}
C. J. Gross et al., Nucl. Instrum. Meth. Section A 450 (2000) 12.

\bibitem{agr16}
J. Agramunt et al., Nucl. Instrum. Meth. Section A 807 (2016) 69.

\bibitem{struck}
http://www.struck.de/

\bibitem{root}
R. Brun and F. Rademakers, Nucl. Instrum. Meth. Section A 389 (1997) 81.

\bibitem{cab17}
R. Caballero-Folch et al., Phys. Rev. C 95 (2017) 064322.

\bibitem{skr74}
K. Skrable, Health Physics 27 (1974) 155.

\bibitem{azu80}
R. E. Azuma et al., Phys. lett. B 96 (1980) 31.

\bibitem{ras18}
B. C. Rasco et al., arXiv:1806.05238; submitted to Nucl. Instrum. Meth. Section A (2018).

\bibitem{tes16}
D. Testov et al., Nucl. Instrum. Meth. Section A 815 (2016) 96.

\bibitem{ensdf}
 Evaluated Nuclear Structure Data File, https://www.nndc.bnl.gov/ensdf/

\bibitem{pho18}
V. H. Phong et al., to be published.

\bibitem{hal18}
O. Hall et al., to be published.

\bibitem{lun80}
E. Lundt et al., Z. Phys. A 294 (1980) 233.

\bibitem{war86}
R. A. Warner and P. L. Reeder, Rad. Effects 94 (1986) 27.

\bibitem{rud93}
G. Rudstam et al., Atom. Data Nucl. Data Tables 53 (1993) 1.

\bibitem{hos10}
P. Hosmer et al., Phys. Rev. C 82 (2010) 025806.

\bibitem{lia18}
J. Liang et al., subm. to Nuclear Data Sheets (2018); 
data available via https://www-nds.iaea.org/relnsd/delayedn/delayedn.html

\bibitem{hua17}
W. J. Huang et al., Chin. Phys. C 41 (2017) 030002; 
M. Wang et al.,  Chin. Phys. C 41 (2017) 030003.

\bibitem{moe03}
P. Moeller et al., Phys. Rev. C 67 (2003) 055802.

\bibitem{kaw08}
T. Kawano et al., Phys. Rev. C 78 (2008) 054601.

\bibitem{hof81}
P. Hoff and B. Fogelberg, Nucl. Phys. A 368 (1981) 210.

\bibitem{kra91}
K.-L. Kratz et al., Z. Phys. A 240 81991) 419.

\bibitem{ver13}
D. Verney et al., Phys. Rev. C 87 (2013) 054307.

\bibitem{got16}
A. Gottardo et al., Phys. Rev. Lett. 116 (2016) 182501.

\end{thebibliography}



\end{document}